\documentclass[11pt, oneside]{amsart}

\usepackage{amsmath}
\usepackage{amssymb}

\usepackage[utf8]{inputenc}
\usepackage[T1]{fontenc}
\usepackage{lmodern}

\usepackage{color}
\usepackage{dcolumn}
\usepackage{float}
\usepackage{graphicx}
\usepackage{multirow}
\usepackage{rotating}
\usepackage{subfigure}
\usepackage{psfrag}
\usepackage{tabularx}
\usepackage[hyphens]{url}
\usepackage{wrapfig}
\usepackage{enumitem}
\usepackage{multicol}
\usepackage{verbatim}

\usepackage[bookmarks, bookmarksopen, bookmarksnumbered]{hyperref}
\usepackage[all]{hypcap}
\definecolor{orange}{rgb}{1.0,0.3,0.0}
\definecolor{violet}{rgb}{0.75,0,1}
\definecolor{darkgreen}{rgb}{0,0.6,0}
\definecolor{cyan}{rgb}{0.2,0.7,0.7}
\definecolor{blueish}{rgb}{0.2,0.2,0.8}

\newcommand{\note}[1]{ {\textcolor{red}    { #1 }}}

\newcommand{\toolname}[1] {\textsf{#1}}

\usepackage[top=2.5cm, bottom=2.5cm, left=2.5cm, right=2.5cm]{geometry}

\sloppypar

\begin{document}

\title[]{Summary of the First Workshop on Sustainable Software for Science: Practice and Experiences (WSSSPE1)}

\author{Daniel~S.~Katz$^{(1)}$,
Sou-Cheng~T.~Choi$^{(2)}$,
Hilmar~Lapp$^{(3)}$,
Ketan~Maheshwari$^{(4)}$,
Frank~L\"{o}ffler$^{(5)}$,
Matthew~Turk$^{(6)}$,
Marcus~D.~Hanwell$^{(7)}$,
Nancy~Wilkins-Diehr$^{(8)}$,
James~Hetherington$^{(9)}$,
James~Howison$^{(10)}$,
Shel~Swenson$^{(11)}$,
Gabrielle~D.~Allen$^{(12)}$,
Anne~C.~Elster$^{(13)}$,
Bruce~Berriman$^{(14)}$
Colin~Venters$^{(15)}$
}

\thanks{{}$^{(1)}$ National Science Foundation, Arlington, VA, USA; Computation Institute, University of Chicago \& Argonne National Laboratory, Chicago, IL, USA; \url{d.katz@ieee.org}}
\thanks{{}$^{(2)}$ NORC at the University of Chicago and   Illinois Institute of Technology, Chicago, IL, USA; \url{sctchoi@uchicago.edu}}
\thanks{{}$^{(3)}$ National Evolutionary Synthesis Center (NESCent),
  Durham, NC, USA; \url{hlapp@nescent.org}}
\thanks{{}$^{(4)}$ Argonne National Laboratory, Chicago, IL, USA; \url{ketan@anl.gov}}
\thanks{{}$^{(5)}$ Center for Computation and Technology, Louisiana State University, Baton Rouge, LA, USA; \url{knarf@cct.lsu.edu}}
\thanks{{}$^{(6)}$ Columbia Astrophysics Laboratory, Columbia University, New
York, NY, USA; \url{matthewturk@gmail.com}}
\thanks{{}$^{(7)}$ Scientific Computing Group, Kitware, Inc.,  Clifton Park, NY, USA; \url{marcus.hanwell@kitware.com}}
\thanks{{}$^{(8)}$ San Diego Supercomputer Center, University of California San Diego, La Jolla, CA, USA, \url{wilkinsn@sdsc.edu}}
\thanks{{}$^{(9)}$ Research Software Development Team, University College London, London, UK; \url {j.hetherington@ucl.ac.uk}}
\thanks{{}$^{(10)}$ University of Texas at Austin, Austin, TX, USA; \url{jhowison@ischool.utexas.edu}}
\thanks{{}$^{(11)}$ University of Southern California, Los Angeles, CA, USA; \url{mdswenso@usc.edu}}
\thanks{{}$^{(12)}$ University of Illinois, Champaign, IL, USA; \url{gdallen@illinois.edu}}
\thanks{{}$^{(13)}$ Norwegian University of Science and Technology, Trondheim, Norway; \url{elster@ntnu.no}}
\thanks{{}$^{(14)}$ Infrared Processing and Analysis Center, California Institute of Technology, Pasadena, CA, USA; \url{gbb@ipac.caltech.edu}}
\thanks{{}$^{(15)}$ University of Huddersfield, Huddersfield, West Yorkshire, UK; \url{C.Venters@hud.ac.uk}}

\begin{abstract} 
Challenges related to development, deployment, and maintenance of 
reusable software for science are becoming a growing concern. 
Many scientists' research increasingly depends  on 
the quality and availability of software upon which their works are built. 
To highlight some of these 
issues and share experiences, the First Workshop on Sustainable 
Software for Science: Practice and Experiences (WSSSPE1) was held 
in November 2013 in conjunction with the SC13 Conference.
The workshop featured keynote presentations and a large number (54)
of solicited extended abstracts 
that were grouped into three themes and presented via panels.
A set of collaborative notes of the presentations and discussion
was taken during the workshop.

Unique perspectives were captured about issues such as comprehensive
documentation, development and deployment practices, software
licenses and career paths for developers. Attribution systems that account for 
evidence of software contribution and impact were also discussed. These include
mechanisms such as Digital Object Identifiers, publication
of ``software papers'', and the use of online systems, for example source code repositories
like GitHub. 

This paper summarizes 
the issues and shared experiences that were discussed, including
cross-cutting issues and use cases. It joins a nascent
literature seeking to understand what drives software work in science,
and how it is impacted by the reward systems of science. 
These incentives can determine the extent to which
developers are motivated to build software for the long-term, for the
use of others, and whether to work collaboratively or separately. It 
also explores community building, leadership, and dynamics in relation to
successful scientific software.


\end{abstract}

\maketitle

\section{Introduction}

The First Workshop on Sustainable Software for Science: Practice and
Experiences (WSSSPE1,
\url{http://wssspe.researchcomputing.org.uk/WSSSPE1}) was held on
Sunday, 17 November~2013, in conjunction with the~2013 International
Conference for High Performance Computing, Networking, Storage and
Analysis (SC13, \url{http://sc13.supercomputing.org}).

Because progress in scientific research is dependent on the quality of
and accessibility to software at all levels, it is now critical to
address many challenges related to the development, deployment, and
maintenance of reusable software.
In addition, it is essential that scientists, researchers, and
students are able to learn and adopt software-related skills and
methodologies. Established researchers are already acquiring some of
these skills, and in particular a specialized class of software
developers is emerging in academic environments as an integral part of
successful research teams. This first WSSSPE workshop provided a forum
for discussion of the challenges around sustaining scientific
software, including contributed short papers in the form of both
positions and experience reports. These short papers, as well as notes
from the debates around them, have been archived to provide a basis
for continued discussion, and have fed into the collaborative writing
of this report.  Some of the workshop submissions have been extended
to full papers, which form part of the same special journal edition in
which this report appears. The workshop generated a high level of
interest; an estimated~90 to~150 participants were in attendance at
different times of the day. The interest and discussions have already
led to follow-up activities: A smaller Python-specific workshop is
planned to be held at the~2014 SciPy conference, and a follow-on
WSSSPE2 workshop has been accepted for
the SC14 conference. In addition, funds to support the workshops have
been obtained from the US National Science Foundation (NSF) and the
Gordon and Betty Moore Foundation, and the original workshop website
has been turned into a community website to engender further
discussion and progress. Additionally, a minisymposium at the 2014
Society for Industrial and Applied Mathematics (SIAM) Annual Meeting
on ``Reliable Computational Science'' (SIAM AN14,
\url{http://meetings.siam.org}) is being co-organized by a WSSSPE1
participant to further explore some of the key issues raised at the
workshop.

This report attempts to summarize the various aspects of the workshop.
The remainder of this paper first gives an overview of the process
with which the workshop was organized (\S\ref{sec:process}), then
proceeds with summaries of the two keynote
presentations~(\S\ref{sec:keynotes}), followed by summaries of the
workshop papers grouped by the three thematic workshop panels to which
they were assigned~(\S\ref{sec:devel}-\ref{sec:community}). Three
broader cross-cutting issues surfaced repeatedly, and are discussed
separately~(\S\ref{sec:cross-cutting}), as are use cases for
sustainable software~(\S\ref{sec:use-cases}). The summaries are based
not only on the papers and panel presentations, but also on the many
comments raised in both the onsite and online discussions that
accompanied the workshop elements, as documented by collaborative note
taking during the workshop~\cite{WSSSPE1-google-notes}. We conclude
with issues and lessons learned, as well as plans for future
activities~(\S\ref{sec:conclusions}).  The original call for papers is included in 
Appendix~\ref{sec:cfp}.  The short papers accepted to
the workshop are listed in Appendix~\ref{sec:papers}, and a partial
list of workshop attendees can be found in
Appendix~\ref{sec:attendees}. 

\section{Workshop Process and Agenda} \label{sec:process}

WSSSPE1 was organized by a relatively small group of five organizers
and a larger program committee of~36 members. The program committee
peer-reviewed submissions, but also had early influence in the
workshop's organization, such as articulating the Call for Papers (see
Appendix~\ref{sec:cfp}).

Aside from setting the stage for the relevance of software
sustainability and corresponding training and workforce issues to
science, the call for papers enumerated the topics it was interested
in as challenges to the ecosystem of which scientific software is a
part, and in which software developers, users, and funders hold
roles. These challenges roughly followed NSF's Vision and Strategy for
Software~\cite{NSF_software_vision}, and specifically included the
development and research process that leads to new (or new versions of
existing) software; the support, community infrastructure, and
engineering for maintenance of existing software; the role of open
source communities and industry; aspects of the use of software, such
as reproducibility, that may be unique to science; policy issues
related to software sustainability such as measuring impact, giving
credit, and incentivizing best practices; and education and training.

The workshop's goal was to encourage a wide range of submissions from
those involved in software practice, ranging from initial thoughts and
partial studies to mature deployments. Consequently, the organizers
aimed to make submission as easy as possible. Rather than requiring
adherence to a formal submission system and a full research
paper-style template, submissions were intentionally limited to short
4-page papers, articulating either a position on one or more of the
topics, or reporting experiences related to them. Furthermore, for
submission authors were asked to self-archive (and thus self-publish)
their papers with a third-party service that issues persistent
identifiers, such as Digital Object Identifiers (DOIs), and to then
submit the URL to the archived paper by email. This had the side
effect that every submitter would also have a publicly available and
citable version of their workshop contribution.

This process resulted in a total of 58 submissions. Almost all
submitters used either arXiv~\cite{arXiv-web} or
Figshare~\cite{figshare-web} to archive their papers. The submissions
were then subjected to peer review by the program committee, resulting
in~181 reviews, an average of~3.12 reviews per paper. Reviews were
completed using a Google form, which allowed reviewers to choose
papers they wanted to review, and to provide general comments as well
as relevance scores to the organizers and to the authors. Based on the
review reports, the organizers decided to list~54 of the papers (see
Appendix~\ref{sec:papers} for a full list) as significant
contributions to the workshop. The high acceptance rate may come as a
surprise, but it is nonetheless consistent with the goal of fostering
broad participation, and as a corollary of the chosen submission
process paper acceptance was no longer a means to filter the papers'
public availability.

Roughly following the call for papers topics, the accepted submissions
were grouped into three main
categories, namely \emph{Developing and Supporting Software},
\emph{Policy}, and \emph{Communities}. Each category was assigned to a
panel, with three to four panelists drawn from authors of the
associated submissions, who were each assigned to read and summarily
present a subset of the papers associated with the panel. The process
from organizing and advertising the workshop, to collecting and
reviewing the papers, and putting together the agenda was documented
by the organizers in a report~\cite{WSSSPE1-pre-report}, which they
self-archived in the same way as contributed papers.

The workshop received submissions from eight North
American and European countries. In some instances authors collaborated across
multiple countries towards jointly authored papers. A majority of
contributions came from the US with 42 papers where at least one author was
affiliated with a US institution. A total of 10 submissions were from 
Europe and 4 were from Canada.  This is not surprising for a workshop
being held in the US.  We believe future versions of the workshop will
have contributions from more countries and more continents.

In terms of subject of the papers, the submissions were dominated by the domain
of practice of sustainable software engineering and management with about 32
papers based on these themes. These papers were further based on a variety of
disciplines including infrastructure and architecture, user engagement, and
governance. Additionally, 18 papers were based on the sciences and
applied mathematics domains with disciplines including High Energy Physics,
Bioinformatics, Nanotechnology, Chemistry, and material sciences. Others were
included topics such as science gateways and visualization.  Again, given that
this workshop was held with a computer and computational science conference,
these numbers are not surprising.

The workshop also included two keynote presentations.
Remote participation was facilitated by a
live-cast of keynotes and panels via Ustream.tv
(\url{http://ustream.tv}).
In each panel, the paper summary presentations were followed by active discussion
involving panelists, onsite attendees, and often online
participants. The latter was facilitated by having a shared Google
Doc~\cite{WSSSPE1-google-notes} for collaborative note taking. Some of
the online discussion also took place on Twitter (hashtag
\texttt{\#wssspe}).

\section{Keynotes } \label{sec:keynotes} 

The WSSSPE1 workshop began with two keynote presentations, which 
resonated with the audience and spurred a number of topics  
discussed throughout the meeting.

\subsection{A Recipe for Sustainable Software, Philip E. Bourne.} \label{sec:keynote1}

The first keynote~\cite{WSSSPE1-keynote1} was delivered by Philip
E. Bourne of University of California, San Diego.  Bourne is a
biomedical scientist who has also formed four software companies. He
co-founded PLOS Computational Biology~\cite{plos-web} and helped
develop the RCSB Protein Data Bank~\cite{pdb-web}.  He is working on
automating three-dimensional visualizations of cell contents and
molecular structures, a problem that has not been solved and when
done, would serve as a key function of software in biomedical
sciences.

Bourne's presentation was based on his own software experiences.  He
emphasized that sustainability for software ``does not just mean more
money from Government'' (see also
Section~\ref{sec:defining-sustainability}).  Other factors to
consider, he mentioned, encompass costs of production, ease of
maintenance, community involvement, and distribution channels.

In places, Bourne said, development in science has improved thanks to
open source and hosting services like GitHub~\cite{github-web}, but
for the most part it remains arcane. He argued that we can learn much
from the App Store model about interfaces, ratings, and so on. He also
mentioned BioJava~\cite{biojava-web} and Open Science Data
Cloud~\cite{osdc-web} as distribution channels.  On a related note,
Bourne observed a common evolutionary pathway for computational
biology projects, from data archive to analytics platform to
educational use, and suggested that use of scientific software for
outreach might be the final step.

Bourne shared with the audience a few real challenges he
encountered. His first anecdote was that he has looked into
reproducibility in computational biology, but has concluded that ``I
have proved I cannot reproduce research from my own
lab''~\cite{Veretnik}.

Another problem Bourne experienced was staff retention from private
organizations which reward those combining research and software
expertise (the ``Google Bus''). However, he is a strong supporter of
software sustainability through public-private partnerships. He noted
that making a successful business from scientific software alone is
rare: founders overvalue while customers undervalue. He noted that to
last, an open source project needs a minimal funding requirement even
with a vibrant community --- goodwill only goes so far if one is being
paid to do something else.  He talked about grant schemes of relevance
in the U.S., particularly with regard to technology
transfer~\cite{sbir-web, fased-web}.

Bourne also had problems with selling research software: the
university technology transfer office wanted huge and unrealistic
intellectual property reach through, whereby they would get a share of
profits from drugs developed by pharmaceutical companies who use the
software.  He advocated for a one-click approach for customers to
purchase university-written software.

He then presented arguments on directly valuing software as a research
output alongside papers, a common discussion within this field.  He
mentioned an exploration of involving software engineers in the review
process of scientific code~\cite{peer-review-code}, and discussed how
publishing software reviews could change attitudes.

On the notion of \emph{digital enterprise}, where information
technology (IT) underpins the whole of organizational activities, he
contended that universities are way behind the curve. In particular,
he highlighted the separation of research, teaching, and
administration into silos without a common IT framework as a blocker
to many useful organizational innovations: ``University 2.0 is yet to
happen.''  He argued that funders such as NSF and NIH can help train
institutions, not just individuals, in this regard.

Bourne concluded by discussing his 2011 paper ``Ten Simple Rules for
Getting Ahead as a Computational Biologist in
Academia''~\cite{bourne_ten} and argued that computational scientists
``have a responsibility to convince their institutions, reviewers, and
communities that software is scholarship, frequently more valuable
than a research article''.

\subsection{Scientific Software and the Open Collaborative Web, Arfon Smith} \label{sec:keynote2}

The second keynote~\cite{WSSSPE1-keynote2} was delivered by Arfon
Smith of GitHub. Smith started with an example from his past in data
reduction in Astronomy, where he needed to remove interfering effects
from the experimental apparatuses. He built a ``bad pixel mask,'' and
realized that while it was persistent, there was no way or practice of
sharing these data among scientists. Consequently many researchers
repeated the same calculations. Smith estimated that plausibly 13 person-years
were wasted by this repetition.

``Why didn't we do better?''  Smith asked of this practice. He argued
this was because we were taught to focus on immediate research
outcomes and not on continuously improving and building on tools for
research. He then asked, when we do know better, why we do not act any
different. He argued that it was due to the lack of incentives: only
the immediate products of research, not the software, are valued.  He
referenced Victoria Stodden's talk at OKCon~\cite{okcon-stodden-talk}
which he said argued these points well.

C. Titus Brown~\cite{ged-web}, a WSSSPE1 contributor, argued that with
regard to reusable software, ``we should just start doing it.''  Smith
replied that documentation should be ``treated as a first class
entity.''  He noted that the open source community has excellent
cultures of code reuse, for example, RubyGems~\cite{rubygems-web},
PyPI~\cite{pypi-web}, and CPAN~\cite{cpan-web}, where there is
effectively low-friction collaboration through the use of
repositories. This has not happened in highly numerical, compiled
language scientific software.  An exception he cited as a good example
of scientific projects using GitHub is the EMCEE Markov Chain Monte
Carlo project~\cite{emcee-web} developed by Dan Foreman-Mackey and contributors.

He argued that GitHub's \emph{Pull Request} code review mechanism
facilitates such collaboration, by allowing one to code first, and
seek review and merge back into the trunk later.

``Open source is \ldots reproducible by necessity,'' Smith quoted
Fernando Perez~\cite{perez-open-src-reproducible}, explaining that
reproducibility is a prerequisite for remote collaboration.  He
pointed out that GitHub could propel the next stage of web
development, i.e., ``the collaborative web,'' following on from the
social web of Facebook.

In conclusion Smith reiterated the importance of establishing
effective incentive models for open contributions and tool builders,
for example, meaningful metrics and research grants such
as~\cite{NSF_software_vision}. He urged computational scientists to
collaborate and share often their research reports, teaching
materials, code, as well as data by attaching proper licenses.

\section{Developing and Supporting Software} \label{sec:devel}

The panel on Developing and Supporting Software examined the
challenges around scientific software development and support, mainly
focused on research groups that in addition to pursuing research also
produce code in various forms. There was widespread agreement that
developing and maintaining software is hard, but best practices can
help. Several participants added that documentation is not just for
users, and writing application programming interface (API)
documentation, tutorials for building and deploying software, together
with documented development practices can be very helpful in bringing
new developers into a project.

Two subjects that prominently surfaced in this panel also came up
throughout other parts of the workshop, and are therefore deferred to
the section on Cross-cutting Issues (\S\ref{sec:cross-cutting}). These
are the lack of long-term career paths for specialists in the various
software development and support areas (see
\S\ref{sec:career-tracks}), and the question of what ``sustainable''
should mean in the context of software (see
\S\ref{sec:defining-sustainability}).

\subsection{Research or Reuse?}

Software is developed for many different purposes, and the
requirements can vary significantly depending on the intended
audience. Most end-users make use of either a graphical user interface
of some kind, or a command line that may offer input and output
formats for running the code and analyzing its output(s). When
discussing backward compatibility it is these various interfaces that
are discussed. For software that builds on other software frameworks
it is the APIs that are most important, and this can encompass issues
such as the source and binary interfaces to the software libraries
developed---with each potentially having a high maintenance cost if
they are to remain compatible over many years. When using command-line
programs it is generally the command-line switches as well as the
input/output formats that could incur significant costs if they are
changed.

There was discussion that backward compatibility is not always
desirable, and it can be very costly. This must be balanced with the
aims of a given project, and how many other projects depend on and use
the code when backwards incompatible changes are to be made. There are
many examples in the wider open source software world of strategies
for dealing with this, and again best practices can go a long way to
mitigating issues around backwards compatibility. Many projects live
with sub-optimal code for a while, and may allow backwards
compatibility to be broken at agreed-upon development points, such as
a major release for a software library.

There were 13 articles about different experiences in this area, but
little about GUI testing, performance, scalability, or agile
development practices. There were several unique perspectives about
issues such as managing API changes, using the same best practices for
software as data, and going beyond simply ``slapping an OSI-approved
license on code.''

It should be noted that several articles that discussed long-term projects, that
could be said to have reached a sustainable period. The Visualization Toolkit
(VTK) was described~\cite{Hanwell_WSSSPE} as being one of the oldest projects
serving as a basis for several other tools such as ParaView~\cite{Hanwell_WSSSPE}, VisIt~\cite{Ahern_WSSSPE}, and VisTrails~\cite{Koop_WSSSPE}. Other examples of
long-term, sustainable projects included MVAPICH~\cite{Panda_WSSSPE} and R/qtl~\cite{Broman_WSSSPE}, which both began development
in 2000, and DUNE~\cite{Blatt_WSSSPE}, which is also over a decade old. In
addition to how long a project has been active, other metrics are important,
such as number of developers, number of institutions, and whether there are
active developers acting as advocates for the continued viability of a project
beyond individual projects and/or institutions.

\subsection{The Importance of Communities}

Communities are extremely important in software projects, and both
their building and continued engagement need attention during the
project life cycle. Several of the submitted papers 
discussed how communities have been built around projects, and what is
needed to enable a project to grow~\cite{Crusoe_WSSSPE, Hanwell_WSSSPE, Christopherson_WSSSPE, Cranston_WSSSPE, Terrel_WSSSPE}. The latter includes public
source-code hosting, mailing lists, documentation, wikis, bug
trackers, software downloads, continuous integration, software quality
dashboards, and of course, a general web presence to tie a project's
channels and artifacts together.

There was extended discussion about the challenge of fostering
communities in which users help each other, rather than always
deferring to the developers of project to answer user
queries. Participants offered examples that this is indeed possible,
such as mailing lists in which developers do not participate much
because users actively respond to questions from other users, but also
asked whether by doing too much the ``core team'' could end up setting
unrealistic expectations. Team Geek~\cite{opac-b1134063} and Turk's
paper on scaling code in the human
dimension~\cite{Turk:2013:SCH:2484762.2484782} discuss how development
lists tend to have many more people contributing when they are
welcoming to people.

\subsection{Software Process, Code Review, Automation, Reproducibility}

The papers submitted to this panel included many general
recommendations for processes, practices, tools, etc. One of the
papers~\cite{Crawford_WSSSPE}
suggested that a ``Software Sustainability
Institute'' should be vested with the authority to develop
standardized quality processes, a common repository, central resources
for services and consulting, a think tank of sorts, and a software
orphanage center (i.e., a place to `take care' of software when the
original developers have stopped doing so). The idea of one common
repository received some resistance, with so many compelling
alternatives available, e.g., Bitbucket or GitHub. The centralized
communication or point of contact was seen as reasonable, with the
statement that ``vested with authority'' is perhaps too strong. However,
``providing tools if needed'' might be more appropriate.

What about actual software engineering principles, such as modularity
and extensibility? This is how industry maintains software, and
ensures it continues to be useful. Often, rewriting software is
considered to be too costly, but with a modular design it can be kept
up to date. Extensibility is expected to keep it relevant, if built
into the project. One counterpoint raised by Jason Riedy was that
trying to take advantage of the latest and greatest hardware often
makes this painful, hence the lack of papers mentioning ``GPUs and
exotic hardware.''

The question of whether funders, such as the NSF, can mandate software
plans in much the same way as they do data management plans, was
raised. Daniel Katz responded that software is supposed to be
described as part of the NSF data management plan, and that in NSF's
definition, data includes software. A comment from Twitter
(@biomickwatson) raised the issue that this requires reviewers and
funders who understand the answers that are given in these
plans. Daniel Katz responded that in programs focused on software or
data this can be done effectively, but agreed that in more general
programs this is indeed a problem.

\subsection{Training Scientists to Develop and Support Software}

Part of the panel discussion focused around community structures and
how academic communities are not taught how to evaluate
cross-disciplinary work. One question raised was whether software
developers can be effective if they are not part of the appropriate
domain community, with responses that this depends on the specific
problem and situation, and that ``T-shaped'' people who have both
disciplinary depth as well as interdisciplinary and collaboration
skills are important~\cite{Trainer_WSSSPE, Dubey_WSSSPE, Girotto_WSSSPE}.
The discussion focused on whether we could
teach software developers and domain scientists to collaborate
together more effectively rather than trying to teach software
developers about domain science and domain scientists about software
development practices. The end goal of this would be to have a single
community with a spectrum of expertise across domain science and
software development, rather than two separate communities~\cite{Prlic_WSSSPE}.

The role of the growing field of team science with software
development was discussed. Team science deals with understanding and
improving collaborative and team-based scientific research, and issues
such as virtual organizations, and tool development across software
development communities~\cite{Christopherson_WSSSPE, Cranston_WSSSPE}.
Further, how should these skills and best practices then be introduced to students?

\subsection{Funding, Sustainability Beyond the First Grant/Institution}

Are there significant differences in projects that have been running
for 1, 3, 5, or 10+ years? Are there shared experiences for projects
of a similar stage of maturity? It was noted that computing and
communication have changed significantly over the past decade, and
many of the experiences are tied to the history of computing and
communication. See the history of GCC, Emacs, or the Visualization
Toolkit for examples. Others felt that computing has changed less, but
communication and the widespread availability of tools has. It was
noted that email lists, websites, chat rooms, version control, virtual
and physical meetings are all over 20 years old.

It appears that while some of the basics of computing may be
similar, the tools commonly used for computing have changed quite
significantly. Reference was made to Perl, which was commonly used,
giving way to whole new languages, such as Python, for gluing things
together and how this induces many students into entirely rewriting
the scaffolding, leaving the old to rot and the experiments to become
non-reproducible as the tools change. There was discussion of this
tendency along with the enormous differences in the speed and ease of
sharing---having to ship tapes around in the early days of software development
(which shaped development of GCC and Emacs in their formative years) as opposed to
the immediate sharing of the latest development online, using revision
control systems like CVS, Subversion, Git, Mercurial, Bazaar, etc.

The question was also posed as to whether the distinction between
researcher and developer is sensible, with James Hetherington
commenting that in the UK a more nuanced view of research software
engineers and researcher developers is examined. Should this be less
of a contract relationship, and more of a collaborative relationship?
This is also at the core of the business model that Kitware presented
in its submission to the workshop. Are other ingredients missing such
as applied mathematicians? Should this be defined more in terms of
skill sets rather than roles and/or identities? This builds on the
comments from Vaidy Sunderam that scientists are generally good
writers, and have mathematical skills, so why can't they learn
software engineering principles?

Miller commented that all of the infrastructure that sits around a new
algorithm that we need to make it useful and sustainable requires
different skill sets than the algorithm developer. Friere commented
that there are no good career paths for people with broad skills, no
incentives for them to continue in these roles. There was debate
around people doing what interests them, and learning computing leaves
people cold, but is it that it leaves the people who find career paths
in academia cold versus the full spectrum of people involved in
research? Is this also caused by poor teaching, or because the
benefits for doing this are perceived as too small? It could also be
attributed to their focus being on science, not software engineering,
or do people with the passion for software engineering in science
simply have no viable career path and either adapt or seek out
alternate career paths?

\section{Policy} \label{sec:policy}

The panel on policy discussed workshop contributions dealing with the
wide range of software sustainability aspects that relate to
establishing, promoting, and implementing policies. Six papers
presented frameworks for defining, modeling, and evaluating software
sustainability, the basis of establishing policies. Four papers
advocated mechanisms for more properly assessing the impact of
scientific software, and for crediting and recognizing work that
promotes software sustainability, all of which are instrumental in
effectively promoting policies that aim to change current
practices. Four papers discuss facets of implementing software
sustainability, and models of implementation across different facets.

\subsection{Modeling Sustainability}

The workshop submissions grouped under this section provide frameworks
for thinking about, researching, and understanding which elements of
sustainability are important and how they are related to each other. Although there is
substantial overlap between the frameworks, they have different
emphases and extents.  Each paper in this group included a definition
of sustainability, with many overlaps between them (see
Table~\ref{tab:defining-sustainability}).  Perhaps unsurprisingly, the
issue of how to define sustainability came to the fore multiple times
during the workshop, and it is thus summarized in depth separately in
~\S\ref{sec:defining-sustainability}.

Table~\ref{tab:defining-sustainability} is based on a summary 
by Lenhardt ~\cite{lenhardt-wssspe1-panel}, which shows for each
contribution to the Modeling Sustainability panel what it meant when
referring to software, how it defined software sustainability, and
which approach it suggested to understand or evaluate sustainability.
\begin{table}[t]
  \begin{scriptsize}
    \begin{center}
      \caption{Summary of Modeling Sustainability papers from Policy Panel.  Adapted from~\cite{lenhardt-wssspe1-panel}.}
      \label{tab:defining-sustainability}
      \begin{tabular}{|p{2.3cm}|p{3.6cm}|p{4.4cm}|p{4.8cm}|}
                \hline
{\bf Paper/Authors}
& {\bf Software}
& {\bf Sustainability}
& {\bf Approach to Understand or Evaluate Sustainability} \\
                \hline
Calero, et al.~\cite{Calero_WSSSPE}
& General notion of software. Not explicitly defined.
& Sustainability is linked to quality.
& Add to ISO \\
                \hline
Venters, et al.~\cite{Venters_WSSSPE}
& Software as science software; increasingly complex; service-oriented computing
& Extensibility, interoperability, maintainability, portability, reusability, scalability, efficiency
& Use various architecture evaluation approaches to assess sustainability \\
                \hline
Pierce, et al.~\cite{Pierce_WSSSPE}
& Cyberinfrastructure software
& Sustainable to the extent to which there is a community to support it
& Open community governance \\
                \hline
Katz, et al.~\cite{Katz_WSSSPE}
& E-research infrastructures (i.e. cyberinfrastructure)
& Persisting over time, meeting original needs and projected needs
& Equates models for the creation of software with sustaining software \\
                \hline
Lenhardt,~et~al.~\cite{Lenhardt_WSSSPE}
& Broadly defined as software supporting science
& Re-use; reproducible science
& Comparing data management life cycle to software development life cycle \\
                \hline
Weber, et al.~\cite{Weber_WSSSPE}
& Software broadly defined; a software ecosystem
& Software niches
& Ecological analysis and ecosystem \\
                \hline
     \end{tabular}
    \end{center}
  \end{scriptsize}
\end{table}

One area in which there was not complete overlap was whether the word
(and thus the effort called for by the WSSSPE workshop) involved
environmental sustainability. Of course the word sustainability has
strong connotations from consideration of environmental issues,
evoking some mention of the areas in which software interacts with
overall environmental resource usage, particularly energy efficiency.
The two papers in this area which mentioned
this~\cite{Venters_WSSSPE,Calero_WSSSPE} did so without integrating
that analysis into the question of software being around long-term,
suggesting that questions of environmental impact of scientific
software is a conceptually distinct area of inquiry.

One group of papers presented frameworks that were primarily about
characteristics of software artifacts, connecting with the long
discourse on software quality. This approach is realized in adjectives
that can be applied to pieces of software but might also extend to
describe software projects.  Thus Calero et al.~\cite{Calero_WSSSPE}
propose adding elements to the ISO standards for measuring software
quality. Specifically, they propose an additional dimension of quality
they call ``perdurability'' with
three defining characteristics: reliability, maintainability, and
adaptability. This overlaps with the framework   by
Venters et al.~\cite{Venters_WSSSPE} who employ the features
``extensibility, interoperability, maintainability, portability,
reusability and scalability,'' anticipating the sorts of work that
people would need to do with a software artifact in the future, ``as
stakeholders requirements, technology and environments evolve and
change.'' Venters et al.~argue that these choices need to be made 
early because they are related to the architecture of the software and
involve trade offs that ought to be analyzed alongside each other. Lenhardt et
al.~\cite{Lenhardt_WSSSPE} compare the software lifecycle to the data
lifecycle to argue for the inclusion of metadata throughout a piece of
software's life (discussing, for example, how it has been built and
tested and what ``data in'' and ``data out'' has been considered). In
addition, their analogy suggests that the software lifecycle might add
a phase of ``preservation'' and draw on the understanding of what that
involves from studies of data.  In sum, then, these frameworks focused
on what needs to be accomplished to have more sustainable software.

A second theme in these papers was the continued availability of
resources to accomplish the goals of sustainability. The elements of
these frameworks focused far more on the organization of a software
project than they did on characteristics of the artifact itself
(although it is certainly true that the adjectives discussed above
could be applied to a software project). For example Pierce et
al.~\cite{Pierce_WSSSPE} focus on the way that the project is run,
particularly in terms of how those involved communicate and jointly
set priorities, a process they call governance. In particular they
argue that because sustainability is related to having ongoing resources,
governance must be open to receive diverse input (by occurring online,
asynchronously, or publicly) and thus have the potential to
``transform passive users into active contributors.'' They argue that
the Apache Software Foundation's incubation process teaches this and
could be learned from by projects throughout scientific software.
Katz and Proctor~\cite{Katz_WSSSPE} also discuss governance,
describing two modes: ``top-down'' and ``bottom-up'' governance.  They
place governance alongside technical questions about the software,
political questions about who is funding
the work surrounding the software, and the
manner in which resources come to the project both initially (commercial,
open source, closed partnerships, grant funded) and long-term (all
four plus paid support).

Frameworks also concerned themselves with the context in which
software projects exist, moving in abstraction from the software
artifact itself to the organization of its production and the shape of
the space in which an artifact or project exists.  These frameworks
take the form of contingency theories in that they outline a different
set of challenges and argue that different project organizations (and
presumably artifact attributes) are necessary to persist long term in
spaces with particular characteristics.  Katz and
Proctor~\cite{Katz_WSSSPE} propose thinking of this idea of space in
terms of three axes: temporal (long or short term needs), spatial
(local or global use) and purpose (specific to general).  They propose
that different project organizations will be needed in different
locations and argue that we should concentrate research to understand
those connections. Weber et al.~\cite{Weber_WSSSPE} describe their
spaces by analogy with natural ecosystems as ``niches'' which sustain
particular pieces of software. They define ``a software niche as the
set of technical requirements, organizational conditions and cultural
mores that support its maintenance and use over time.'' They call for
better understanding and modeling of niches (as well as further
exploration of the usefulness of ecosystem metaphors).

\subsection{Credit, Citation, and Impact}

How work on scientific software is recognized and rewarded strongly
influences the motivation for particular kinds of work on scientific
software. A recurring theme of the panel discussion was that software
work in science is inadequately visible within the reputation system
underlying science; in other words it often doesn't ``count''. In his
paper for this workshop, Katz placed software work along with other
``activities that facilitate science but are not currently rewarded or
recognized''~\cite{Katz2_WSSSPE}. Priem and Piwowar argued for the need
to ``support all researchers in presenting meaningful impact evidence
in tenure, promotion, and funding applications.''~\cite{Priem_WSSSPE}.
Knepley et al.~argued that the lack of visibility of software that
supported a piece of science ``can have detrimental effects on
funding, future library development, and even scientific
careers.''~\cite{Knepley_WSSSPE}.

These papers, and the discussion at the workshop, join a nascent
literature seeking to understand what drives software work in science
and how the reward systems of science thereby shape the type of
software work undertaken. This study includes the extent to which developers are
motivated to build software for the long-term, for the use of others,
and whether to work collaboratively or
separately~\cite{howison_incentives_2013, howison_scientific_2011,
  bietz_synergizing_2010}. Software work is not only motivated by
direct citations, but the visibility of software work in the
literature is important to those who write software used in science.

Papers and discussion concentrated on three overarching questions: How
ought software work be visible, what are the barriers to its
visibility, and what can be done to make it more visible?

Most of the papers in this area focused on visibility of software in
scientific papers, since scientific papers are the most widely
accepted documentation of achievement in science. It was noted that
there appear to be no widely accepted standards on how the use of
software towards a paper ought to be mentioned, and that journals,
citation style guides and other guides to scientific conduct are vague
about how to describe software. To address this, papers advocated the
need for a fixed identifier for software, either directly through a
mechanism such as a Digital Object
Identifier~\cite{Katz2_WSSSPE,Knepley_WSSSPE} or via a published paper
written to document the software (and perhaps its creation), a
``software paper''~\cite{Chue_Hong_WSSSPE}. However, as was pointed
out during the panel discussion, one of the problems with papers as
the cited product is that their author list is fixed in time, which
discourages potential contributors who are not on the original author
list from designing incremental improvements as integration work
rather than separate (and hence possibly rewritten) software
products~\cite{howison_incentives_2013}.

Another approach is to reduce the difficulty of citing all software
underlying a research paper. For example, scientists often work with
software that itself wraps other software, leading to attribution
stacking that can make it non-obvious or even difficult to determine
what attributions would be appropriate. Knepley et
al.~\cite{Knepley_WSSSPE} approach this by proposing a mechanism by
which the software itself, after it has run, provides the user with a
set of citations, according to the pieces of code actually
executed. They describe a prototype implementation whereby the
citations are embedded in libraries and reported along with the
results, via a command-line
interface~\cite{Knepley_WSSSPE}. Discussion highlighted the difficulty
that attempting to acknowledge the contributions of all pieces of
dependent code within a paper faces the difficulty of creating very
long citation lists, straining the analogy of code used to papers
cited. Katz approaches this issue by proposing a system of
``transitive credit,'' recording dependencies and relative
contributions outside particular papers, relieving authors from the
responsibility of acknowledging each and every dependency. Instead
authors would acknowledge the percentage contribution of the software
they used directly and an external system would then be able to
recursively allocate that credit to those who had provided
dependencies~\cite{Katz2_WSSSPE}. Finally Priem and Piwowar argued that machine learning
techniques could examine the body of published literature and extract
mentions of software, coping with the multitude of informal ways in
which authors mention software they used~\cite{Priem_WSSSPE}.  A point
raised in the panel discussion was that instead of asking users to
improve their software citation practices, one can also ask how
software projects can better monitor the literature to improve their
ability to show impact. For example, the nanoHUB project scans the
literature using keywords and names of known users to discover papers
that are likely to have used their software and platform, and assigns
graduate students to read each paper, highlighting mentions of
software use and sometimes following up with the authors to identify
stories for demonstrating impact.  A process for tracking
software-using publications with the goal of increasing impact
visibility is now described at \url{http://publications.wikia.com}.

Potential visibility, and thus acknowledgement of scientific software
products is not restricted to publications. Another key location for
visibility is in the grant funding process, and as emphasized by NSF
representatives at the meeting, recent changes to grant proposal and
reporting formats now allow both applicants and awardees to report and
highlight software products as much as publications. Nonetheless,
whether peer review panels would value these contributions in the same
way as publications remains to be seen.

Priem and Piwowar argued that assessing the impact of software work
requires looking beyond publications, including evidence of
contribution and impact recorded in social coding-oriented resources
such as GitHub, and conversations about software in issue trackers,
mailing lists, twitter and beyond~\cite{Priem_WSSSPE}. In keeping with
a principle of the ``altmetrics'' approach, they advocate that
scholars should have resources that empower them to tell their own
stories in the manner most appropriate for them and their audiences.

\subsection{Implementing Policy}

The workshop contributions in this group were concerned with the
aspect of how implementation of best practices and other
recommendations for improving scientific software sustainability could
be promoted. Specifically, if scientific software is to become more
sustainable, corresponding policies and guidelines need to be such
that the scientific community can follow and implement them. This is
considerably more challenging that it might seem at first, because in
the reality of science today resources, both financial and personnel,
that could be devoted to implementation are very limited, and the
reward system does not encourage scientists to do so. Furthermore,
implementing sustainability-targeting policies and guidelines often
takes a variety of specialized software engineering expertises, which
are not necessarily found in a single engineer, and much less so in a
domain scientist cross-trained in programming. Adding to the policy
implementation challenges, applicable sustainability-promoting
practices and guidelines will change through a software project's
lifecycle, in particular as it gains maturity.

Two of the papers in this group focus on specific facets of software
design that are important factors in a project's sustainability but
are often addressed only late in the scientific software development
cycle, if at all: Krintz et al.~\cite{Krintz_WSSSPE} look at API
governance, and Heiland et al.~\cite{Heiland_WSSSPE} discuss maturity
models for software security. The other two papers discuss
implementation strategies for science from the perspective of
facilitating many or all facets of sustainability-oriented software
design: Blanton and Lenhardt~\cite{Blanton_WSSSPE} contrast large
projects that have software infrastructure development built-in, with
cross-training domain scientist PIs in software engineering best
practices. Huang and Lapp~\cite{Huang_WSSSPE} discuss how various
specialized software engineering skills could be turned into shared
instrumentation with low barriers to access.

Krintz et al.~\cite{Krintz_WSSSPE} describe how in an era in which
computing frequently takes place in a distributed cloud, the control
over digital resources is increasingly shifting from physical
infrastructure to APIs, in particular web-service APIs. Yet, as Krintz
et al.~observe, unlike for physical IT infrastructure in data centers,
science communities have developed very little in the way of practices
and technology for API governance, referred to by Krintz et al.~as the
``combined policy, implementation, and deployment control''. Web APIs
can and do change, sometimes quite frequently, raising the need to
port dependent applications. The effort required for porting is
notoriously difficult to estimate, making it nearly impossible for IT
organizations to assess and thus properly manage the impact of API
changes. To address this, Krintz et al.~propose a mechanism that
evaluates the porting effort between two versions of a web-service API
in a formal and automated way. To analyze the porting effort, they
divide API changes into \emph{syntactic similarity}, the changes in
inputs and outputs, and into \emph{semantic similarity}, the changes
in the API's behavior. In initial tests, their method showed good
congruence with human developers in scoring porting effort, offering
the possibility that API governance can become as solid a part of
scientific IT management as data center infrastructure management is
today.

Many facets of engineering for software sustainability strongly depend
on the maturity level of the software. However, the maturity level of
a software project meant to be sustained changes tremendously over its
development life cycle, and the eventual maturity level is often
difficult to predict during initial development. Using software
security as their case study, Heiland et al.~\cite{Heiland_WSSSPE}
discuss how Maturity Models can be used to formalize best practices
appropriate for the different levels of maturity that a software
project may go through over its lifecycle. Cybersecurity is also an
example of a sustainability-relevant aspect in software design that is
rarely given due diligence in science. In particular in industry,
cybersecurity best practices for different stages of life cycle and
maturity have been formalized as Software Security Maturity Models,
and are widely used, yet awareness of these among scientific software
development communities remains low. In providing a path to tightening
security practices as software matures, such models align with the
objective of providing implementation approaches that the scientific
community can actually follow.

API governance and cybersecurity measures appropriate for a project's
maturity are all but two facets of sustainability-oriented software
development. Others include user-centered interface design, test
engineering, dependency management, and deployment operations. Each of
these facets requires specialized skills and training in software
engineering. How can the implementation of best practices and
guidelines along many or all of these different facets be facilitated
in scientific software development? Blanton and
Lenhardt~\cite{Blanton_WSSSPE} contrast two models. In one, the time
and personnel devoted to software engineering is ``co-funded'' with
the driving science project. This typically implies large multi-year
collaborative projects that to succeed require significant software
infrastructure to be built, and which thus have the funding to support
one or several software engineers. The sustainability of such projects
then depends on sustaining the funding. In the other model, domain
scientists also take the role of software developers, whether by
necessity such as funding limitations, or due to cross-disciplinary
professional interests. For this to result in sustainable software
products, the domain scientists need to be (or become) cross-trained
in software engineering standards and best practices.

In practice, there are example for both extremes of the spectrum. For
example, in the life sciences the iPlant Collaborative~\cite{iPlant},
the Galaxy Project~\cite{Galaxy}, and \toolname{Qiime}~\cite{Qiime}
are large multi-year projects with significant software infrastructure
funding and deliverables. Typically though these are the exception
rather than the rule, and particularly in the long tail of science the
scientist-developer predominates, even for software that is widely
used or crosses domain boundaries such as
\toolname{rOpenSci}~\cite{rOpenSci}. For better training domain
scientists at least in basic software engineering best practices,
initiatives such as Software Carpentry~\cite{SoftwareCarpentry} have
demonstrated how this could be achieved at scale.

However, there may also be a middle ground between the two
extremes. Huang and Lapp~\cite{Huang_WSSSPE} propose a Center of
Excellence model that leverages economies of scale to make software
engineering experts and their skills accessible to the long tail of
science. As Huang and Lapp discuss, this model could effectively turn
the utilization of software engineering expertise from a complex human
resource recruitment and management challenge, to buying time, when
and to the extent needed, on shared instrumentation. There is
precedence to using such a model to lower access barriers for long
tail science, in particular for new experimental technologies.  For
example, although the acquisition and operation of next-generation
high-throughput DNA sequencers requires substantial investments of
money, time, and expertise, the establishment of ``core facilities''
on many university campuses has made these technologies accessible to
a wide swath of scientists, with transformative results for science.

One of the important conclusions from this group of papers is that
creating sustainable software requires paying attention not to one or
two, but to several different facets of software engineering, each
with corresponding best practices and standards of excellence. Even if
a science project requires and has funding for a full software
engineering FTE, what the project really needs could be fractions of
different engineers with different specialty training. The vast
majority of long tail science projects lacks the funding for even one
full software engineer, let alone one who combines expertise in all of
the applicable facets of engineering. Some scientists in the long tail
will have the professional interests to cross-train enough in software
engineering to be successful with creating sustainable software, but
it is unrealistic to expect this expertise and interest of all or even
the majority. This is where a software engineering center of
excellence could provide a critical resource by enabling scientists in
the long tail to complement their resources and expertise with facets
that are missing, but which, if applied at the right time, would
improve the chances of a software product to become sustainable. Such
complementary expertise also need not be restricted to software
engineering in the strict sense; for example, it could consist of
community building, leadership, and support for some period of
transition to sustainability.

In summary, implementing software sustainability practices on a
broader basis requires on the one hand the development of guidelines
and practices that are suitable for the exploratory research context
in which most scientific software is created, and on the other hand a
skilled workforce trained in a variety of software engineering facets
and community building. The capabilities afforded by such a workforce
need to be accessible not only to large projects with sufficient
funding to provide competitive employment, but also to the many
smaller projects in the long tail of science. Sufficiently
cross-training domain scientists could be one way to achieve this;
another, complementary approach is to instrument the necessary
capabilities so they can be shared.

\section{Communities} \label{sec:community}

Across all talks and papers submitted, authors implicitly and
explicitly recognized the concept of ``communities'' as a driving and
unifying force within software projects.  Despite that, the actual
nature of the communities, the incentive structures that bind them
together, the infrastructure they utilize for communication and even
the types of individuals that make up those communities were
different.  Some of these communities were primarily composed of
members of industry, some were funded and driven by individuals
focused on developing software as opposed to utilizing it, and others
were primarily composed of scientist practitioners, and were true
communities of practice.

These varying structures and compositions result in differing modes of
interaction within communities, styles of development, and the
structure of planning for future development of software.  In this
section, we summarize the different types of communities in scientific
software as well as the resultant impact on sustainability and
development of functionality.

\subsection{Communities}


Drawing on experiences from high-energy physics, Vay et
al.~\cite{Vay_WSSSPE} proposed developing teams of technical
specialists   to overcome a lack of coordination between projects.
Maximizing scientific output requires maximizing the usability of the
scientific code while minimizing the cost of developing and
supporting those code.  This included targeting different
architectures for their software to be deployed, as well as
coordination between technically-focused individuals and usage of a
common scripting language between projects.  Instead of fragmenting
the development of simulation codes across institutions, the paper
suggests that a cohesive strategy reducing duplication and increasing
coordination will broadly increase the efficiency across institutions.
The approach proposed is of de-fragmenting the existing ecosystem in a
non-disruptive way.

Maheshwari et al.~\cite{Maheshwari_WSSSPE} focuses on ``technology
catalysts'' and their role in the modern scientific research
process. A technology catalyst is an individual with knowledge of
technological advancements, tasked with user engagement to create
scientific or engineering applications, using suitable tools and
techniques to take advantage of current technological capabilities.
One of the tasks of technology catalysts is to seek community
collaborations for new applications and engage users, thus benefiting
both science, by effective running of scientific codes on
computational infrastructure, and technology, by conducting research
and seeking findings for technology improvement.  The particular
engagements described in the paper came up from the lead author's work
as a postdoctoral researcher at Cornell and Argonne, where interaction
with the scientific communities in both institutions resulted in these
collaborations.

At NESCent, a combination of in-house informatics individuals and
domain scientists collaborate to develop software to study
evolutionary science.  The report~\cite{Cranston_WSSSPE} studied the
success of a ``hackathon'' model for development, where short-form,
hands-on events combining users, researcher-developers and software
engineers targeted specific code improvements. From this experiment,
the authors identified several key outcomes as well as
lessons-learned: specifically, the co-localization of developers was
seen as having a strong impact, enabling casual conversation that led
to discrete outcomes.  The formation of the discussion mailing list,
specifically in response to the social capital built at the hackathon,
was seen as spurring on longer-term benefits to the community and
fostering sustainability.

Hart et al.~\cite{Hart_WSSSPE} addresses the success of the rOpenSci
project in developing collaboration supporting tools for
Open Science.  This software collective, organized around the
statistical programming environment R, develops access mechanisms for
data repositories and attempts to reduce the barrier to entry for
individuals wanting to access data repositories and study the data
contained therein.  The collective fosters direct collaboration
between individuals and data providers, designed to ``train academics
in reproducible science workflows focused around R.''  Two central
challenges are engagement of existing users within ecology and
evolutionary biology (EEB), and how the community can make inroads and
traction in other disciplines. Currently, the collective is exploring
addressing these challenges through the use of social media, holding
workshops and hackathons.  This helps to both raise the profile of
the collective within EEB and in other domains.  However, the
overarching challenge identified in the paper was that of
incentivizing maintenance of software, which is difficult in academia.



Christopherson et al.~\cite{Christopherson_WSSSPE} outlines the degree
to which research relies on high quality software. There are often
barriers and a lack of suitable incentives for researchers to embrace
software engineering principles. The Water Science Software Institute
is working to lower some of these barriers through an Open Community
Engagement Process. This is a four-step iterative development process
that incorporates Agile development principles.

\begin{itemize}
\setlength{\itemindent}{0.5in}
\item[Step 1:] Design - thorough discussion of research questions
\item[Step 2:] Develop working code
\item[Step 3:] Refine based on new requirements
\item[Step 4:] Publish open source
\end{itemize}

Christopherson reports on the application of Steps~1-3 to a
computational modeling framework developed in the 1990s. Step~1 was a
2-day, in-person specifications meeting and code walk-through. Step 2
was a 5-day hackathon to develop working code, and Step 3 was a 3-day
hackathon to refine the code based on new requirements. The team
worked on small, low-risk units of code. It was challenging, revealed
unanticipated obstacles, programmers had to work together, and
experimentation was encouraged.
The paper recommended: start small and gradually building toward more
complex objectives, consistent with Agile development; develop
consensus before coding, by repeating step~1 before all hackathons;
ensure newcomers receive orientation prior to the hackathon, such as a
code walk-through or system documentation; and co-locate collaborators
whenever feasible.


Pierce et al.~\cite{Pierce2_WSSSPE} describes how science gateways can
provide a user-friendly entry to complex cyberinfrastructure. For
example, more than 7,000 biologists have run phylogenetic codes on
supercomputers using the CIPRES Science Gateway in $3 \tfrac{1}{2}$
years. Over 120 scientists from 50 institutions used the
\toolname{UltraScan Science Gateway} in one year, increasing the
sophistication of analytical ultracentrifugation experiments
worldwide. The new Neuroscience Gateway (\toolname{NSG}) registered
more than 100 users who used 250,000 CPU hours in only a few months.

Gateways, however, need to keep operational costs low and can often
make use of common components, such as authentication, application
installation and reliable execution and help desk support. Science
Gateway Platform as a Service (\toolname{SciGaP}) delivers middleware
as a hosted, scalable third-party service while domain developers
focus on user interfaces and domain-specific data in the gateway.
While \toolname{SciGaP} is based on the \toolname{Apache Airavata}
project and the \toolname{CIPRES} Workbench Framework, community
contributions are encouraged by its open source, open governance and
open operations policies. The goal is robust, sustainable
infrastructure with a cycle of development that improves reliability
and prioritizes stakeholder requirements. The project is leveraging
Internet2's \toolname{Net+} mechanisms for converting
\toolname{SciGaP} and its gateways into commodity services.

Zentner et al.~\cite{Zentner_WSSSPE} describes experiences and
challenges with the large \toolname{nanoHUB.org} community, where
community is defined as a ``body of persons of common and especially
professional interests scattered through a larger society.'' Support
is challenging because of the diversity of viewpoints and needs. The
group constantly examines its policies to determine whether they are
indirectly alienating part of the community or encouraging particular
types of use.
\toolname{nanoHUB}'s 10-year history with over 260,000 users annually
provides a lot of data to analyze: 4000 resources contributed by 1000
authors. \toolname{nanoHUB} serves both the research and education
community and the contribution model allows researchers to get their
codes out into the community and in use in education very rapidly. The
primary software challenges are twofold --- support for the
\toolname{HUBzero} framework and challenges related to the software
contributed by the community.

The group has learned that community contributions are maximized with
a tolerant licensing approach. \toolname{HUBzero} uses an LGPLv3
license so contributors can create unique components and license as
they choose. If they make changes to source code, the original license
must be maintained for redistribution. As far as contributed
resources, these must be open access, but not necessarily open
source. This allows contributors to meet the requirements of their
institutions and funding agencies. Quality is maintained via user
ratings. Documentation is encouraged and nanoHUB supplies regression
test capabilities, but the user community provides ratings, poses
questions and contributes to wishlists and citation counts, all of
which incentivize code authors.

Terrel~\cite{Terrel_WSSSPE} describes support for the Python
scientific community through two major efforts: the SciPy conference
and the NumFOCUS foundation.  Since software sustainability relies on
contributions from all sectors of the user community, these efforts
support these sectors, and help develop and mature Python.
The reliance on software in science has driven a huge demand for
development, but this development is typically done as a side effort
and often in a rush to publish without documentation and
testing. While the software is often created by academics, software
support can fall to industrial institutions. SciPy brings together
industry, government, and academics to share their code and experience
in an open environment.
NumFOCUS in a non-profit that promotes open, usable scientific
software while sustaining the community through educational programs;
collaborative research tools and documentation; and promotion of
high-level languages, reproducible scientific research, and open-code
development. Governance is a loose grantor-grantee relationship with
projects, allowing funds to be placed in the groups' accounts. This
has raised money to hire developers for open code development,
maintain testing machines, organize the PyData conference series, and
sponsor community members to attend conferences.

L\"{o}ffler et al.~\cite{Loffler_WSSSPE} describes the Cactus project,
which was started in 1996 by participants in the USA Binary Black Hole
Alliance Challenge. Cactus has a flesh (core) and thorns (modules)
model, a community-oriented framework that allows researchers to
easily work together with reusable and extendable software
elements. Modules are compiled into an executable and can remain
dormant, becoming active only when parameters and simulation data
dictate. Users can use modules written by others or can write their
own modules without changing other code. The community has grown and
diversified beyond the original science areas.

The paper points out four keys to sustaining the community: modular
design, growing a collaborative community, career paths, and credit. In
modular design, the Cactus project went far beyond standard practices
of APIs. Domain specific languages (DSLs) allow decoupling of
components --- for example I/O, mesh refinement, PAPI counters, and
boundary conditions abstracted from science code. In academia,
publications are the main currency of credit. Because the project
connects code developments to science, the work is publishable and
modules are citable. Because of the open source, modular approach,
programmers can see the impact of their contributions and often
continue work after graduation. Career paths remain a challenge,
however. Tasks that are essential from a software engineering
perspective are often not rewarded in academia. The best programmers
in a science environment often have multidisciplinary expertise. This
also is not rewarded in academia.

Wilkins-Diehr et al.~\cite{Wilkins-Diehr_WSSSPE} describes an NSF
software institute effort to build a community of those creating
science gateways for science. These gateways, as described in some of
the other papers in this section, can be quite capable and can have
strong scientific impact.  Challenges are similar to those highlighted
by other papers in this section: the conflict between funding for
research vs infrastructure and the challenges around getting academic
credit for infrastructure.
The authors observe that development is often done in an isolated
hobbyist environment. Developers are unable to take advantage of
similar work done by others, finding themselves in isolation even when their projects have common
goals. But often projects struggle for sustainable funding because
they provide infrastructure to conduct research and many times only
the research is funded. Gateways also may start as a small group
research project, taking off in popularity once they go live, without
any long term plans for sustainability.  Subsequent disruptions in
service can limit effectiveness and test the limits of the research
community's trust. 

Recommendations from an early study of successful gateways include: 1)
Leadership and management teams should design governance to represent
multiple strengths and perspectives, plan for change and turnover in
the future, recruit a development team that understands both the
technical and domain-related issues, consider sustainability and
measure success early and often. 2) Projects should hire a team of
professionals, demonstrate credibility through stability and clarity
of purpose, leverage the work of others, and plan for flexibility. 3)
Projects should identify one or more existing communities and understand the
communities' needs before beginning, then adapt as the communities'
needs evolve. 4) Funders should consider the technology project
lifecycle, and design solicitations to reward effective planning,
recognize the benefits and limitations of both technology innovation
and reuse, expect adjustments during the production process, copy
effective models from other industries and sectors, and encourage
partnerships that support gateway sustainability.


\subsubsection{What are communities?}

The workshop did not directly answer the question ``What are
communities?'' but instead a number of different answers were
indirectly presented, through the depiction of individuals and
stakeholders in different aspects of the scientific software
lifecycle.  In broad strokes, however, scientific software communities
were generally accepted as consisting of individuals, often but not
always composed of scientist practitioners, that were working with
some degree of coordination toward a common goal enabled by software.

The discussion of development-focused communities centered around
describing methods of interaction between individuals and the
scientific software.  The first type of interaction was the
development of a specific piece of software, the second was a
particular domain or discipline, and the final primary type of
interaction was around the development of applications built on a
particular piece of software that was perhaps developed by another
group.  As an example, in~\cite{Hanwell_WSSSPE}, the community
described is comprised of both the for-profit company
\toolname{Kitware} and the users and contributors to their software
packages such as \toolname{VTK}.  This structure, of the centralized
development of core infrastructure around which communities of
individuals applying that infrastructure and developing applications
utilizing it, was similarly reflected in~\cite{Terrel_WSSSPE}, where
the core scientific python ecosystem is supported by a non-profit
entity that fosters community investment in that ecosystem.  In many
ways, these two organizations (\toolname{Kitware} and
\toolname{NumFOCUS}) attempt to cross domain boundaries and provide
support for both the infrastructure and application sides of community
building.

\subsubsection{Measuring community}
How might a project know when it has built a sustainable community?
How might an outsider be able to assess the activity and
sustainability of that community?  These questions have been partially
addressed in the literature.  For example, a key metric in online
communities in general is the cross-over point where there are more
external contributors than internal ones. Richard Millington's book
``Buzzing Communities'' does an excellent job of outlining these
measures, drawing on communities research in an accessible
manner~\cite{millington_buzzing_2012}. Some participants in the
workshop have since prepared materials outlining current and
future practices in measurement of scientific software and its
ecosystem~\cite{metrics-web}.

\subsubsection{Additional Resources for learning about software communities}

Scientific software communities were viewed as a subset of software
communities as whole.  As such, resources applicable to generic
software communities -- such as open source and proprietary technology
companies -- can be used as input and as guiding understanding of how
to steward and develop scientific software communities.  Because
incentive structures are different in industry and volunteer-based
open source communities, these can provide rough guidelines but not
necessarily identically applicable.  The analogy between corporations
and scientific investigators (particularly in terms of competition,
cooperation and competitive advantage) has been explored in the
literature below, but because of the different incentive structure the
analogy is not universally true.

The literature below, suggested by attendees, addresses both
non-scientific software projects, as well as scientific projects.  The
selections address both descriptive and prescriptive approaches to
communities.

Both~\cite{howison_scientific_2011} and~\cite{howison_incentives_2013}
study how scientific software collectives self-organize and address
issues of incentive, long-term support, and development of
infrastructure as well as new features.  As noted elsewhere in this
summary, \cite{Turk:2013:SCH:2484762.2484782} shared prescriptions
from two software communities in astrophysics.

From the perspective of developing prescriptions for successful
scientific software development, both~\cite{citeulike:11831265} and
\cite{1749-4699-6-1-015010} share experiences and suggestions for
developing sustainable practices.  \cite{citeulike:11831265} proposes
"ten simple rules" for developing open source scientific software,
focusing on both the choices made during development and the
sustainability of practices in the long term.
\cite{1749-4699-6-1-015010} describes the development and long-term
growth of the \texttt{deal.II} library, and how its place in its
ecosystem of libraries, applications and domains has shaped its
development and community trajectory.

From more traditional open source development, resources were shared
that developed communities explicitly, such
as~\cite{citeulike:7888211} and~\cite{Trapani:2011}, focusing on
large-scale projects such as the Ubuntu Linux distribution and
smaller-scale volunteer-developed projects like such as ThinkUp,
respectively. The process of open source development, while less
explicitly focused on community building, sketched
in~\cite{citeulike:478633} was seen as a valuable resource,
particularly when combined with the management and personal
interaction techniques outlined in~\cite{opac-b1134063}.  Growing
diversity in communities was directly addressed
in~\cite{Allsopp:2012}, where experiences growing the diversity of
technical conferences in open source were described.

\subsection{Industry \& Economic Models}

Several papers presented discussed the connection between industry and
scientific software, from the perspective of both integrating efforts
between the two and sustaining long-term development.

Hanwell et al.~\cite{Hanwell_WSSSPE} reflect on the 15-year history of
open source software development at \toolname{Kitware}.  In
particular, they focus on their success at growing their community of
users through enabling multiple channels of communication, directly
reaching out to individuals, and lowering the barrier to entry for
contributions. This involves providing clear, test-oriented and
review-based mechanisms for evaluating contributions, permissive
licenses, and a service-based model for sustaining development.  This
model enables \toolname{Kitware} to receive both public funding, as
well as private funding to support improvements and targeted
developments of the software.

Foster et al.~\cite{Foster_WSSSPE} discussed the approach of
developing sustainability models around Software-as-a-Service (SaaS)
platforms, with the target example being that of
\toolname{GlobusOnline}.  The authors build a case that both
grant-based and volunteer-based development fall short in sustaining
software, resulting in software that is disproportionately difficult
to use compared to its functionality, which they note directly impacts
the overall scientific productivity of its users.  In contrast, by
charging a subscription fee for hosted, centrally-managed software
(similar to offerings by \toolname{Dropbox}, \toolname{EverNote},
\toolname{GMail}), the authors propose to manage the funding cycle and
enable a greater focus on the aspects of software that directly impact
individuals, rather than funders.  \toolname{Globus} has deployed such
a service, for which they have attempted to develop a sustainable
economic model that reduces institutional obstacles to funding and
subscription.  However, they do identify that cultural obstacles do
still remain, and they note a particular difference in culture between
NSF- and NIH-funded researchers.

\subsection{Education \& Training}

The papers describing education and training were focused primarily on
how these aspects of community development impact on the long-term
sustainability of software projects.  \cite{Girotto_WSSSPE} described
the impact of the mandate within the International Centre for
Theoretical Physics (ICTP) to foster resources and competences in
software development and HPC, resulting in the development of
educational curricula directed in this area.  The paper itself
described the changes made in these curricula as a result of the
current changes in the HPC and scientific software landscape due to
the advent of scripting languages, new programming paradigms and new
types of hardware such as accelerator technologies.  The development
of a workshop, with carefully selected participants and an immersive
approach to learning, was identified as a major success for educating
and developing new scientific software developers from targeted
domains. 

Elster~\cite{Elster_WSSSPE} also points out how the 
prevalence and rapid growth of multi and many-core systems forces 
awareness of data locality and synchronization issues
if one want to teach people how to develop high-performing scientific codes.

\cite{Crawford_WSSSPE} addressed education and training within
computational chemistry frameworks, particularly as these frameworks
attempt to address next-generation computer hardware and software and
as chemistry courses emphasize lab work over computational education.
The authors identify this lack of computational awareness and training
as the primary challenge to future advances in computational
chemistry.  The authors propose a new institute for computational
chemistry, emphasizing collaboration (and a licensing structure, such
as LGPL or more permissive) and education of future generations of
chemistry researchers.

%
%
%

\section{Cross-cutting Issues} \label{sec:cross-cutting}

Three issues, how to define software sustainability, how career
paths (or their lack) interact with achieving it, and the impact of
software licenses, were raised across
the workshop's panels.  This section aims to synthesize these
discussions from different perspectives.

\subsection{Defining Sustainability}  \label{sec:defining-sustainability}

What is, or should be meant by ``sustainability'' in the context of
software came up in many different parts of the workshop, specifically
in the first keynote (\S\ref{sec:keynote1}), the Developing and
Maintaining Software panel (\S\ref{sec:devel}), and the Policy panel
(\S\ref{sec:policy}). It quickly became clear that at present there is
no consensus among the community, whether within or across
disciplines, on what this definition should be, and that a variety of
different definitions were being assumed, used, or sometime expressly
articulated by contributors and attendees. However, some concepts,
particularly relating sustainability to change over time, were also
evidently held in common. This common notion is, for example, captured
in the definition used by the UK's Scientific Software Sustainability
Institute, quoted in~\cite{Venters_WSSSPE}: ``software you use today
will be available---and continue to be improved and supported---in the
future''. Pierce et al~\cite{Pierce_WSSSPE} express this idea as
software that continues to serve its users.

Philip Bourne, too, used the relation to change over time when he
suggested in his opening keynote that sustainability can perhaps be
defined as the effort needed to make the essential things continue.
This leads to having to decide what it is that we want to sustain,
whether what we want to sustain is valuable, and finally, who would
care and how much if it went away. As was pointed out during a
discussion session, OSS Watch, an advisory service for issues relating
to free and open source software in the UK, proposes a Software
Sustainability Maturity Model to address the issue of what level of
sustainability a particular element of software needs to have, and
where this is important. It, too, expresses sustainability in relation
to change over time:
\begin{quote}``When choosing software for procurement or
development reuse --- regardless of the license and development model
you will use --- you need to consider the future. While a software
product may satisfy today's needs, will it satisfy tomorrow's needs?
Will the supplier still be around in five years' time? Will the
supplier still care for all its customers in five years' time? Will
the supplier be responsive to bug reports and feature requests? In
other words, is the software sustainable?''~\cite{OSS-ssmm-web}
\end{quote}

Attendees suggested that having a definition of sustainability on
which the community can agree is key.  A related question that was
raised is what the goal of sustainability should be, with a wide range
of possible answers, including more reproducible science, software
persistence, and quality.  And given a goal of sustainability, how
would success in achieving it be measured?  How would one know that a
piece of software has reached sustainability? Participants pointed out
that for truly sustainable software there should be no endpoint at
which sustainability can be claimed, because the software products
would continue to be used and useful beyond the single institution,
grant, and developer or development team that created them. This may
mean that sustainability needs to be addressed throughout the full
software life cycle.  It was also pointed out that software
sustainability is not isolated from other attributes of scientific
software and its use, such as usability, and provenance. Similarly,
the question was considered, albeit only briefly, whether proprietary
versus open-source license plays a role in the context of software
sustainability. For example, should a project ensure that it uses an
OSI-approved license so that software products can outlive any single
entity if they remain important.

Because part of the Policy panel (\S\ref{sec:policy}) was about
modeling sustainability, and modeling requires defining what will be
modeled, this panel saw particular attention to the questions
surrounding the definition of sustainability. Two papers, Venters et
al.~\cite{Venters_WSSSPE} and Katz and Proctor~\cite{Katz_WSSSPE},
specifically discuss the issue.

According to Venters et al.~\cite{Venters_WSSSPE}, sustainability is a
rather ambiguous concept, and the lack of an accepted definition
hampers integrating the concept into software engineering. They
suggest that sustainability falls under the category of non-functional
requirements, and that a software's sustainability is a consequence of
a set of central software architecture quality attributes, including
extensibility, interoperability, maintainability, portability,
reusability, and scalability. They also propose an evaluation
framework with which quality and sustainability could be measured at
the architectural level.

Katz and Proctor~\cite{Katz_WSSSPE} propose a set of questions that
could be used to measure software sustainability:
\begin{itemize}
\item Will the software continue to provide the same functionality in
  the future, even if the environment (other parts of the
  infrastructure) changes?
\item Is the functionality and usability clearly explained to new
  users?
\item Do users have a mechanism to ask questions and to learn about
  the software?
\item Will the functionality be correct in the future, even if the
  environment changes?
\item Does it provide the functionality that existing and future users
  want?
\item Will it incorporate new science, theory, or tools as they
  develop?
\end{itemize}
Despite their phrasing, these questions are not intended to be given
simplistic yes or no answers, and it is the complete set rather than
any individual one that would determine where in the range of
sustainability a software falls.

\subsection{Career Paths for Scientific Software Developers}\label{sec:career-tracks} 

Career path issues also came up repeatedly, starting in the first
keynote (\S\ref{sec:keynote1}), where Phil Bourne used the term ``the
Google Bus'' to describe the issue of well-trained software
development staff in academic labs choosing to leave science and to
work instead for technology firms, especially Google, which happens in
large enough numbers that Google operates a bus every day to its
nearest offices (and hence staff who leaves academia in this way do
not even have to physically move).

The career path issue emerged repeatedly across panels because for
scientific software to be(come) sustainable, projects trying to create
sustainable software need to be able to recruit and retain software
developers trained in the various requisite software engineering
facets.  However, a career path in research means faculty at most
universities, and as was noted repeatedly in discussions, faculty
are hired based on their scientific qualifications, not on their
software development skills or track record. Consequently, developing
special software development skills is unlikely to further a career in
science at a university, although national laboratories were
acknowledged as a different case. Loffler et
al.~\cite{Loffler_WSSSPE}, one of the papers in the Communities panel
(\S\ref{sec:community}), brought the problem to the point:
\begin{quote}
  ``The most severe problem for developers in most computational
  sciences currently is that while most of the work is done creating
  hopefully well-written, sustainable software, the academic success
  is often exclusively tied to the solution of the scientific problem
  the software was designed for. Tasks that from a software
  engineering standpoint are essential, e.g., high usability,
  well-written and updated documentation, or porting infrastructure to
  new platforms, are not rewarded within this system.'' ~\cite{Loffler_WSSSPE}
\end{quote}

Clearly, improving the recognition of software engineering work is
connected to addressing the career path problem. As was noted in the
Developing and Supporting Software (\S\ref{sec:devel}) and the Policy
(\S\ref{sec:policy}) panel discussions, there are encouraging signs of
improvement, including some altmetric services (such as Impactstory,
\url{http://impactstory.org}) collecting metrics for software source
code, and the fact that NSF now asks to list ``products'' rather than
their ``publications'' in an investigator's biosketch or results from
prior NSF support. However, how software, let alone parts of software
are reused by others can be very difficult to measure, and better
recognizing software products for principle investigators by itself
does not create career paths for specialist software developers
working as part of a university research group. Huang and
Lapp~\cite{Huang_WSSSPE}, a contribution to the Policy panel, offer
one possible solution in which a software engineering center of
excellence offers a career path for a correspondingly trained
workforce, and increased recognition of the resulting more sustainable
software would in a virtuous cycle heighten the value of the center's
services.

\subsection{Licensing and Software Patents}
Issues related to licensing and patents primarily was discussed in the Communities
session, but licensing was also a concern of many of the other contributors
in other sessions.
Software is no longer just open or closed (only binaries available),
but also licensed and patented, which clearly also impacts 
software sustainability.
While many papers briefly discussed licensing issues,
Elster~\cite{Elster_WSSSPE} directly discussed the impact of software licenses 
on obtaining industrial funding for scientific software
projects.  In particular, she described her experiences with researchers
unwilling to utilize GPL (copyleft) code, as it adds restrictions to
reuse that they themselves as well as some industries find unacceptable for future
commercialization.  (This was discussed by Hanwell et
al.~\cite{Hanwell_WSSSPE} as well.)

US information technology companies funding 
academics will thus often insist on BSD licensing on software 
since they then can legally include the code into their commercial
codes.
On the other hand, there are companies
that fund larger GPL-licensed software projects~\cite{Blatt_WSSSPE}
and insist that the university projects they fund 
also produce code with GPL licensing.
They do not accept BSD-like licenses since they argue that
other companies then may choose to build closed commercial
codes on they software they funded, rather than encouraging 
the community to contribute freely and thus ensuring software 
sustainability for the community.
In either case, the university researchers are not given much choice
if they want these much sought after funds in a increasingly competitive
grant world.

Another obstacle to
sustainability identified by Elster include patenting of software.
Most countries place some limits on software patents. The European 
Union outright forbids them, while US patent law excludes ``abstract ideas'', 
which has been used to refuse some software patents.
Further obstacles to
sustainability include a lack of open access, and even more broadly, a lack of open source
codes even in open access journals.  Finally, a lack of awareness on
the part of scientific software developers of commodity libraries for
common tasks reduces their ability to reuse code.

\section{Case Studies} \label{sec:use-cases} 

In this section, we discuss some of the software projects as case
studies to better understand the points discussed during the workshop
and described in the previous sections, and to find how 
they are affected by sustainability issues in
practice. 
%
Most of the software projects discussed here were publicly launched 10
or more years ago. We generally note the original release (o.r.) year of each project in 
parenthesis in its first mention.

We classify the software projects discussed in the workshop in two
broad categories. First, the \emph{utility} software, comprising of
general purpose software. Utility software is often used as enabler
or facilitator for the development of other tools and techniques
to carry out scientific work. This includes the software developed to
efficiently utilize new research infrastructures. Second, the
\emph{scientific} software, comprising the software that was
originally developed with an aim to solve a specific scientific
problem.
%
This classification is motivated by our argument that the two kinds of software
projects wildly vary in factors such as scope, purpose and usage. The
development and management of each kind is significantly different.
Consequently, the sustainability challenges faced by them differ and must be
treated separately. For instance, the challenges faced by a gateway software
development project such as \toolname{CIPRES} (o.r. 2007) or visualization software
products such as \toolname{VisIT (o.r. 2001)} or \toolname{ParaView} (o.r.  2002) are
distinct to a niche software for \textit{ab initio} modeling and simulation
such as \toolname{VASP} (o.r. 1992) or \toolname{Quantum Espresso} (o.r. 2001). 

\subsection{Utility Software}
Software developed with a potentially wider audience and general purpose usage
in mind is utility software. Utility software typically does not address
fundamental research problems for a given scientific domain. Examples are
collaborative development frameworks such as \toolname{GitHub} (o.r. 2008) and
\toolname{Bitbucket} (o.r.  2008), distributed workflow and generic computing
frameworks such as \toolname{Galaxy} (o.r. 2006), \toolname{HUBzero} (o.r. 2010),
\toolname{SimGrid} (o.r. 2001), \toolname{Swift} (o.r. 2007), \toolname{Globus} (o.r. 2000) and
\toolname{VisTrails} (o.r. 2007), and visualization frameworks such as \toolname{VTK},
\toolname{VisIT}, and \toolname{ParaView}.

Development is often a high risk/reward undertaking exploring
uncharted territories and is largely influenced by (re)usability
factors. Owing to a relatively large number of features, the development
and prototype process is also lengthy which poses a significant
survival risk. Challenges on a class of utility software for new
architectures is well discussed in~\cite{Ferenbaugh_WSSSPE}.

On the other hand, utility software presents opportunities to be usable by a
larger community making its undertaking and development an attractive pursuit.
It is generally more visible in community which in turn leads to a broader and
deeper participation. For instance, it helps promoting collaborations across
the breadth (e.g., different departments) and depth (e.g., stakeholders within
a department) of community, one of the key ingredients of a sustainable
process. Successful utility projects reap high rewards and have a longer usage
span. Development process becomes user-driven and self-sustaining.

One such example is the \toolname{Galaxy} project~\cite{Galaxy}. It follows
agile software development practices and implements standards such as
test-driven development and socialized bug managing practices via
\toolname{trello}. \toolname{Galaxy} \emph{histories} and \emph{toolshed} offer
easy community sharing of data and tools further promoting a collaborative
environment. The project closely follows the guidelines described in Carver and
Thiruvathukal~\cite{Carver_WSSSPE} and many from Prli\'{c} and
Procter~\cite{Prlic_WSSSPE}. Many utility software projects are often developed
aiming better utilizing a particular, new infrastructure and architecture,
e.g., \toolname{MVAPICH} (o.r. 2002), \toolname{VisIT}, \toolname{ParaView}. Similarly, to
leverage the power of emerging architectures such as accelerators, new code and
libraries are required. The experience of one such effort as described in
Ferenbaugh~\cite{Ferenbaugh_WSSSPE} which met with a limited success but
nonetheless with many invaluable lessons were learned about influential
cultural and technical aspects in sustainable software development practices.

A relatively new paradigm in utility software is the software
delivered as service over the web. With increasing popularity of
cloud-based storage and computational environments, many users are
leaning towards tools used as services. \toolname{GitHub} and
\toolname{Bitbucket} can be argued to be such tools, catering to
collaborative development. For scientific users
\toolname{Globus}-based tools are a case of service-based utility
software discussed during the workshop. The data movement and sharing
services offered by \toolname{Globus} can be easily used over the web by
collaborating researchers.

\subsection{Scientific software}
Scientific software consists primarily of special-purpose software
that was purpose-built for a target use-case scenario, fixed 
requirements in mind, or solving a specific problem. Software projects
pertaining to specific scientific domain tend to be in a niche and the
user community tends to be small to medium. They are mostly driven by
the science and specific needs of a research group. Specific needs
such as numerical accuracy and algorithmic optimization are some of
the paramount requirements of most scientific software.

Long-term sustainability of scientific software is often a significant
challenge and face radically different issues compared to utility
software.  Many submissions reported that software is practically
considered a ``byproduct'' of the actual research. Others contended
that the software was not the main funded part of their research. A
smaller codebase and fixed requirements result in stability, ease of
installation, and configuration. Many such projects mature and are
treated as libraries to be integrated into larger systems such as some of
the utility software discussed in the previous section. While the
software can stay stable and require relatively low maintenance, the
responsibility is often on the shoulders of a few developers who
are often not specialists in software development. Development tends to
be linear and simplistic with a limited scope to follow software best
practices.

Some examples of such software discussed as part of the workshop are
\toolname{DUNE (o.r. 2008)}, \toolname{R/qtl} (o.r. 2002), \toolname{Kitware} (o.r. 1998),
\toolname{PETSc} (o.r. 1995), and \toolname{MINRES-QLP} (o.r. 2007),
most of which are focused on one scientific or applied
mathematics domain. However, sometimes such projects grow beyond the
initial vision of developers. One such example is \toolname{Kitware},
which while being a software product specializing in the scientific
process, has a core focus of developing communities around software
processes. Another instance of this process is the development of the
\toolname{CMake} build utility, which started out as a building tool
for \toolname{ITK} but grew to become a generic build utility for C++
projects. Similarly, \toolname{PETSc} is growing towards becoming a
general purpose utility system usable for solving a variety of
scientific problems.


\subsection{Distinctions}

In conclusion, we find that there are distinctions in the
characteristics and challenges faced between utility and
scientific software projects.  
We 
find that the
utility software packages are more likely to use the best
practices discussed during the workshop. Often, sustainability of
scientific software projects is achieved by the fact that the core
developer or team heavily utilizes the software for their own science,
e.g., \toolname{R/qtl, PETSc}.  Furthermore, the development of
scientific software requires more scientific background compared with
utility software, thus in many cases, the bulk of development
is carried out by a domain scientist. For these reasons, we
believe that separate guidelines and sustainability principles could
be defined for these two software categories.


\section{Conclusions} \label{sec:conclusions}

To conclude, we highlight what we have learned from the workshop, and what we plan to do going forward.



\subsection{Issues and lessons}


Three major issues came up repeatedly in different parts of the workshop:

\begin{enumerate}
\item The need for a definition of sustainability such that the
  community can get behind it. Although some had hoped that at least
  an initial consensus could be reached in the course of the workshop,
  this proved elusive. However, in the absence of such a definition it
  will remain difficult to define exactly what the goals should be
  towards achieving, or even only improving software sustainability,
  and hence what practices should be followed and implemented when. As
  described in the next subsection (\S\ref{sec:future}), the workshop
  organizers have begun an effort to address this.

\item The need for academic career paths for scientific software
  developers.  Unfortunately, it is not clear how to ensure that these
  career paths become available, other than repeatedly talking about
  this issue. The recent Moore and Sloan initiative in data
  science~\cite{moore_sloan} are trying to address this, to some
  extent, by providing funds and incentives to universities in the US
  that work towards this goal.

%
%
%

\item The need for recognition of scholarship in scientific software
  over research articles. This need probably is the most addressed of
  the three, today, with efforts underway such as the Mozilla Science
  Lab, GitHub, and figshare ``Code as a research object''
  project~\cite{code_as_a_research_object} among others.

\end{enumerate}
In addition, licensing and patents, and how they impact research funding
for software development, were also discussed.


\subsubsection*{Discussion sessions}

Two strong lessons came out of the three discussion sessions:

\begin{enumerate}
\item Use of shared repositories in the development of collaborative
  projects facilitates collaboration, reproducibility, and
  sustainability in computational science. However, it represents a
  barrier in some scientific fields and has yet to be more widely
  adopted.




\item A sustainability model for scientific software is to build a
  pipeline from construction to consumption, as found in the most
  efficient information technology enterprises.

\end{enumerate}


\subsubsection*{Use cases}
Two distinct class of scientific software projects and products were
discussed in the workshop: 1) generic, large-scale utility software
and 2) niche, medium- and small-scale scientific software. Each class
faces different and significant challenges. New undertakings should
recognize the differences in advance and identify such challenges
within the development and sustaining efforts. In particular, the
dynamics associated with developers, scope, life cycle,
users-community, (re)usability, funding support, and career paths vary
widely among the two classes of software.


\subsubsection*{Workshop process}
The WSSSPE workshop can be viewed as an experiment in how we can
collaboratively and inclusively build a workshop agenda, without
asking a large number of people to submit papers that will be rejected
so that the workshop can have a low acceptance rate.

Contributors also want to get credit for their participation in the
process.  And the workshop organizers want to make sure that the
workshop content and their efforts are recorded.  The methods used in
the WSSSPE1 workshop were successful: we had good participation;  
contributors have a report they can cite;   the record of the
workshop is open and available through the self-published reports,
the workshop website and notes site, and this paper.  In addition,
many additional papers are being created that will include the
discussions at the workshop, including extended versions of many of
the self-published reports such as those that are in this special
issue.

Ideally, there would be a service that would be able to index the
contributions to the workshop, serving the authors, the organizers,
and the larger community.

\subsection{Future activities} \label{sec:future}

The organizers of the workshop have begun a survey to understand how
the community define software sustainability.  It is expected that
this survey will gather one or more consensus definitions, and lead to
a short paper discussing them, as well as the level of consensus.

Additional activities that are being planned include two additional
WSSSPE workshops at the 2014 SciPy and SC14 conferences.  The SciPy
workshop (WSSSPE1.1) will focus on how some software projects from the
SciPy community have dealt with software sustainability issues, both
successfully and unsuccessfully, while the SC14 workshop (WSSSPE2)
will be more general, and will likely focus on determining and
publicizing specific activities that the overall scientific software
community can take to move forward.  In addition, there will be a
two-session minisymposium on ``Reliable Computational Science'' at the 2014
Society for Industrial and Applied Mathematics Annual Meeting (SIAM
AN14, \url{http://meetings.siam.org}) to further explore some of the
key issues raised here.

\section*{Acknowledgments}


Some of the work by Katz was supported by the National Science
Foundation while working at the Foundation; any opinion, finding, and
conclusions or recommendations expressed in this material are those of
the author and do not necessarily reflect the views of the National
Science Foundation.

Choi thanks Ian Foster, Director of the Computation Institute,
University of Chicago, for encouraging her work in reliable
reproducible research and supporting her trip to participate in the
WSSSPE1 conference as a contributor.

Lapp was supported by the National Evolutionary Synthesis Center
(NESCent), NSF EF-0905606.

Hetherington was supported by the UK Engineering and Physical Sciences
Research Council (EPSRC) Grant EP/H043160/1 for the UK Software
Sustainability Institute.

Howison was supported by NSF SBE-1064209 and NSF ACI-0943168.

Wilkins-Diehr was supported by the Extreme Science and Engineering Discovery Environment (XSEDE), 
NSF ACI-1053575.

Elster would like to thank Statoil ASA for supporting her travel
and her research group through a grant related to the OPM/DUNE open source projects.

\bibliographystyle{vancouver}

\newpage

\bibliography{wssspe_paper}

\newpage

\appendix

\section{Call For Papers} \label{sec:cfp}

The full Call for Papers for WSSSPE1 can be found online at
\url{http://wssspe.researchcomputing.org.uk/wssspe1/cfp/}. We include
below the part that explains scope and topics for which submissions
were solicited (see \S\ref{sec:process} for number of submissions
received, accepted, and the corresponding process).

\begin{quote}
``Progress in scientific research is dependent on the quality and
accessibility of software at all levels and it is now critical to
address many new challenges related to the development, deployment,
and maintenance of reusable software. In addition, it is essential
that scientists, researchers, and students are able to learn and adopt
a new set of software-related skills and methodologies. Established
researchers are already acquiring some of these skills, and in
particular a specialized class of software developers is emerging in
academic environments who are an integral and embedded part of
successful research teams. This workshop will provide a forum for
discussion of the challenges, including both positions and
experiences. The short papers and discussion will be archived as a
basis for continued discussion, and we intend the workshop to feed
into the collaborative writing of one or more journal publications.

In practice, scientific software activities are part of an ecosystem
where key roles are held by developers, users, and funders. All three
groups supply resources to the ecosystem, as well as requirements that
bound it. Roughly following the example of NSF's Vision and Strategy
for Software~\cite{NSF_software_vision},
the ecosystem may be viewed as having challenges related to:

\begin{itemize}[leftmargin=0.2in]
\item the development process that leads to new (versions of) software
\begin{itemize}[leftmargin=0.2in]
\item how fundamental research in computer science or
  science/engineering domains is turned into reusable software
\item software created as a by-product of research
\item impact of computer science research on the development of
    scientific software and vice versa
\end{itemize}
\item the support and maintenance of existing software, including
  software engineering
\begin{itemize}[leftmargin=0.2in]
\item governance, business, and sustainability models
\item the role of community software repositories, their operation and
  sustainability
\end{itemize}
\item the role of open source communities or industry
\item use of the software
\begin{itemize}[leftmargin=0.2in]
\item growing communities
\item reproducibility, transparency needs that may be unique to science
\end{itemize}
\item policy issues, such as
\begin{itemize}[leftmargin=0.2in]
\item measuring usage and impact
\item software credit, attribution, incentive, and reward
\item career paths for developers and institutional roles
\item issues related to multiple organizations and multiple countries,
  such as intellectual property, licensing, etc.
\item mechanisms and venues for publishing software, and the role of
  publishers
\end{itemize}
\item education and training
\end{itemize}

This workshop is interested in all of the above topics.  We invite
short (4-page) position/experience reports that will be used to
organize panel and discussion sessions. These papers will be archived
by a third-party service, and provided DOIs [Digital Object
  Identifiers].  We encourage submitters to license their papers under
a Creative Commons license that encourages sharing and remixing, as we
will combine ideas (with attribution) into the outcomes of the
workshop.  An interactive site will be created to link these papers
and the workshop discussion, with options for later comments and
contributions. Contributions will be peer-reviewed for relevance and
originality before the links are added to the workshop site;
contributions will also be used to determine discussion topics and
panelists. We will also plan one or more papers to be collaboratively
developed by the contributors, based on the panels and discussions.''
\end{quote}

\newpage

\section{Papers Accepted and Discussed at WSSSPE1} \label{sec:papers}

\subsection*{Developing and Supporting Software}

\subsubsection*{Development Experiences}

\begin{itemize}

\item Mark C. Miller, Lori Diachin, Satish Balay, Lois Curfman
  McInnes, Barry Smith. Package Management Practices Essential for
  Interoperability: Lessons Learned and Strategies Developed for
  FASTMath~\cite{Miller_WSSSPE}

\item Karl W. Broman, Thirteen years of R/qtl: Just barely sustainable~\cite{Broman_WSSSPE}

\item Charles R. Ferenbaugh, Experiments in Sustainable Software
  Practices for Future Architectures~\cite{Ferenbaugh_WSSSPE}

\item Eric G Stephan, Todd O Elsethagen, Kerstin Kleese van Dam, Laura
  Riihimaki. What Comes First, the OWL or the Bean?~\cite{Stephan_WSSSPE}

\item Derek R. Gaston, John Peterson, Cody J. Permann, David Andrs,
  Andrew E. Slaughter, Jason M. Miller, Continuous Integration for
  Concurrent Computational Framework and Application Development~\cite{Gaston_WSSSPE}

\item Anshu Dubey, B. Van Straalen. Experiences from Software
  Engineering of Large Scale AMR Multiphysics Code Frameworks~\cite{Dubey_WSSSPE}

\item Markus Blatt. DUNE as an Example of Sustainable Open Source
  Scientific Software Development~\cite{Blatt_WSSSPE}

\item David Koop, Juliana Freiere, Cl\'{a}udio T. Silva, Enabling
  Reproducible Science with VisTrails~\cite{Koop_WSSSPE}

\item Sean Ahern, Eric Brugger, Brad Whitlock, Jeremy S. Meredith,
  Kathleen Biagas, Mark C. Miller, Hank Childs, VisIt: Experiences
  with Sustainable Software~\cite{Ahern_WSSSPE}

\item Sou-Cheng (Terrya) Choi. MINRES-QLP Pack and Reliable
  Reproducible Research via Staunch Scientific Software~\cite{Choi_WSSSPE}

\item Michael Crusoe, C. Titus Brown. Walking the talk: adopting and
  adapting sustainable scientific software development processes in a
  small biology lab~\cite{Crusoe_WSSSPE}

\item Dhabaleswar K. Panda, Karen Tomko, Karl Schulz, Amitava Majumdar.
The MVAPICH Project: Evolution and Sustainability of an Open Source
Production Quality MPI Library for HPC~\cite{Panda_WSSSPE}

\item Eric M. Heien, Todd L. Miller, Becky Gietzel, Louise
  H. Kellogg. Experiences with Automated Build and Test for
  Geodynamics Simulation Codes~\cite{Heien_WSSSPE}

\end{itemize}

\subsubsection*{Deployment, Support, and Maintenance of Existing Software}

\begin{itemize}

\item Henri Casanova, Arnaud Giersch, Arnaud Legrand, Martin Quinson,
  Fr\'{e}d\'{e}ric Suter. SimGrid: a Sustained Effort for the
  Versatile Simulation of Large Scale Distributed
  Systems~\cite{Casanova_WSSSPE}

\item Erik Trainer, Chalalai Chaihirunkarn, James Herbsleb. The Big
  Effects of Short-term Efforts: A Catalyst for Community Engagement
  in Scientific Software~\cite{Trainer_WSSSPE}

\item Jeremy Cohen, Chris Cantwell, Neil Chue Hong, David Moxey,
  Malcolm Illingworth, Andrew Turner, John Darlington, Spencer
  Sherwin. Simplifying the Development, Use and Sustainability of HPC
  Software~\cite{Cohen_WSSSPE}

\item Jaroslaw Slawinski, Vaidy Sunderam. Towards Semi-Automatic
  Deployment of Scientific and Engineering Applications~\cite{Slawinski_WSSSPE}

\end{itemize}

\subsubsection*{Best Practices, Challenges, and Recommendations}

\begin{itemize}

\item Andreas Prli\'{c}, James B. Procter. Ten Simple Rules for the
  Open Development of Scientific Software~\cite{Prlic_WSSSPE}

\item Anshu Dubey, S. Brandt, R. Brower, M. Giles, P. Hovland,
  D. Q. Lamb, F. Löffler, B. Norris, B. O'Shea, C. Rebbi, M. Snir,
  R. Thakur, Software Abstractions and Methodologies for HPC
  Simulation Codes on Future Architectures~\cite{Dubey2_WSSSPE}

\item Jeffrey Carver, George K. Thiruvathukal. Software Engineering
  Need Not Be Difficult~\cite{Carver_WSSSPE}

\item Craig A. Stewart, Julie Wernert, Eric A. Wernert, William
  K. Barnett, Von Welch. Initial Findings from a Study of Best
  Practices and Models for Cyberinfrastructure Software Sustainability~\cite{Stewart_WSSSPE}

\item Jed Brown, Matthew Knepley, Barry Smith. Run-time extensibility:
  anything less is unsustainable~\cite{Brown_WSSSPE}

\item Shel Swenson, Yogesh Simmhan, Viktor Prasanna, Manish Parashar,
  Jason Riedy, David Bader, Richard Vuduc. Sustainable Software
  Development for Next-Gen Sequencing (NGS) Bioinformatics on Emerging
  Platforms~\cite{Swenson_WSSSPE}

\end{itemize}

\subsection*{Policy}

\subsubsection*{Modeling Sustainability}

\begin{itemize}

\item Coral Calero, M. Angeles Moraga, Manuel F. Bertoa. Towards a
  Software Product Sustainability Model~\cite{Calero_WSSSPE}

\item Colin C. Venters, Lydia Lau, Michael K. Griffiths, Violeta
  Holmes, Rupert R. Ward, Jie Xu. The Blind Men and the Elephant:
  Towards a Software Sustainability Architectural Evaluation Framework~\cite{Venters_WSSSPE}

\item Marlon Pierce, Suresh Marru, Chris Mattmann. Sustainable
  Cyberinfrastructure Software Through Open Governance~\cite{Pierce_WSSSPE}

\item Daniel S. Katz, David Proctor. A Framework for Discussing
  e-Research Infrastructure Sustainability~\cite{Katz_WSSSPE}

\item Christopher Lenhardt, Stanley Ahalt, Brian Blanton, Laura
  Christopherson, Ray Idaszak. Data Management Lifecycle and Software
  Lifecycle Management in the Context of Conducting Science~\cite{Lenhardt_WSSSPE}

\item Nicholas Weber, Andrea Thomer, Michael Twidale. Niche Modeling:
  Ecological Metaphors for Sustainable Software in Science~\cite{Weber_WSSSPE}

\end{itemize}

\subsubsection*{Credit, Citation, Impact}

\begin{itemize}

\item Matthew Knepley, Jed Brown, Lois Curfman McInnes, Barry
  Smith. Accurately Citing Software and Algorithms Used in
  Publications~\cite{Knepley_WSSSPE}

\item Jason Priem, Heather Piwowar. Toward a comprehensive impact
  report for every software project~\cite{Priem_WSSSPE}

\item Daniel S. Katz. Citation and Attribution of Digital Products:
  Social and Technological Concerns~\cite{Katz2_WSSSPE}

\item Neil Chue Hong, Brian Hole, Samuel Moore. Software Papers:
  improving the reusability and sustainability of scientific software~\cite{Chue_Hong_WSSSPE}

\end{itemize}

In addition, the following paper from another area were also discussed
in this area.

\begin{itemize}

\item Frank L\"{o}ffler, Steven R. Brandt, Gabrielle Allen and Erik
  Schnetter. Cactus: Issues for Sustainable Simulation Software~\cite{Loffler_WSSSPE}

\end{itemize}

\subsubsection*{Reproducibility}

\begin{itemize}

\item Sou-Cheng (Terrya) Choi. MINRES-QLP Pack and Reliable
  Reproducible Research via Staunch Scientific Software~\cite{Choi_WSSSPE}

\item Victoria Stodden, Sheila Miguez. Best Practices for
  Computational Science: Software Infrastructure and Environments for
  Reproducible and Extensible Research~\cite{Stodden_WSSSPE}

\end{itemize}

\subsubsection*{Implementing Policy}

\begin{itemize}

\item Randy Heiland, Betsy Thomas, Von Welch, Craig Jackson. Toward a
  Research Software Security Maturity Model~\cite{Heiland_WSSSPE}

\item Brian Blanton, Chris Lenhardt, A User Perspective on Sustainable
  Scientific Software~\cite{Blanton_WSSSPE}

\item Daisie Huang, Hilmar Lapp. Software Engineering as
  Instrumentation for the Long Tail of Scientific Software~\cite{Huang_WSSSPE}

\item Chandra Krintz, Hiranya Jayathilaka, Stratos
  Dimopoulos, Alexander Pucher, Rich Wolski. Developing Systems for API Governance~\cite{Krintz_WSSSPE}

\end{itemize}

\subsection*{Communities, Models, and Education}

\subsubsection*{Communities}

\begin{itemize}

\item Reagan Moore. Extensible Generic Data Management Software~\cite{Moore_WSSSPE}

\item Karen Cranston, Todd Vision, Brian O'Meara, Hilmar Lapp. A
  grassroots approach to software sustainability~\cite{Cranston_WSSSPE}

\item J.-L. Vay, C. G. R. Geddes, A. Koniges, A. Friedman,
  D. P. Grote, D. L. Bruhwiler. White Paper on DOE-HEP Accelerator
  Modeling Science Activities~\cite{Vay_WSSSPE}

\item Ketan Maheshwari, David Kelly, Scott J. Krieder, Justin
  M. Wozniak, Daniel S. Katz, Mei Zhi-Gang, Mainak
  Mookherjee. Reusability in Science: From Initial User Engagement to
  Dissemination of Results~\cite{Maheshwari_WSSSPE}

\item Edmund Hart, Carl Boettiger, Karthik Ram, Scott
  Chamberlain. rOpenSci -- a collaborative effort to develop R-based
  tools for facilitating Open Science~\cite{Hart_WSSSPE}

\item L. Christopherson, R. Idaszak, S. Ahalt. Developing Scientific
  Software through the Open Community Engagement
  Process~\cite{Christopherson_WSSSPE}

\item Marlon Pierce, Suresh Marru, Mark A. Miller, Amit Majumdar,
  Borries Demeler. Science Gateway Operational Sustainability:
  Adopting a Platform-as-a-Service Approach~\cite{Pierce2_WSSSPE}

\item Lynn Zentner, Michael Zentner, Victoria Farnsworth, Michael
  McLennan, Krishna Madhavan, and Gerhard Klimeck, nanoHUB.org:
  Experiences and Challenges in Software Sustainability for a Large
  Scientific Community~\cite{Zentner_WSSSPE}

\item Andy Terrel. Sustaining the Python Scientific Software
  Community~\cite{Terrel_WSSSPE}

\item Frank L\"{o}ffler, Steven R. Brandt, Gabrielle Allen and Erik
  Schnetter. Cactus: Issues for Sustainable Simulation
  Software~\cite{Loffler_WSSSPE}

\item Nancy Wilkins-Diehr, Katherine Lawrence, Linda Hayden, Marlon
  Pierce, Suresh Marru, Michael McLennan, Michael Zentner, Rion
  Dooley, Dan Stanzione. Science Gateways and the Importance of
  Sustainability~\cite{Wilkins-Diehr_WSSSPE}

\item Sou-Cheng (Terrya) Choi. MINRES-QLP Pack and Reliable
  Reproducible Research via Staunch Scientific Software~\cite{Choi_WSSSPE}

\end{itemize}

In addition, the following paper from another area was also discussed
in this area.

\begin{itemize}

\item Marcus Hanwell, Amitha Perera, Wes Turner, Patrick O'Leary,
  Katie Osterdahl, Bill Hoffman, Will Schroeder. Sustainable Software
  Ecosystems for Open Science~\cite{Hanwell_WSSSPE}

\end{itemize}

\subsubsection*{Industry \& Economic Models}

\begin{itemize}

\item Anne C. Elster. Software for Science: Some Personal
  Reflections on Funding, Licensing, Publishing and Teaching~\cite{Elster_WSSSPE}

\item Ian Foster, Vas Vasiliadis, Steven Tuecke. Software as a Service
  as a path to software sustainability~\cite{Foster_WSSSPE}

\item Marcus Hanwell, Amitha Perera, Wes Turner, Patrick O'Leary,
  Katie Osterdahl, Bill Hoffman, Will Schroeder. Sustainable Software
  Ecosystems for Open Science~\cite{Hanwell_WSSSPE}

\end{itemize}

In addition, the following papers from other areas were also discussed
in this area.

\begin{itemize}

\item Brian Blanton, Chris Lenhardt, A User Perspective on Sustainable
  Scientific Software~\cite{Blanton_WSSSPE}

\item Markus Blatt. DUNE as an Example of Sustainable Open Source
  Scientific Software Development~\cite{Blatt_WSSSPE}

\item Dhabaleswar K. Panda, Karen Tomko, Karl Schulz, Amitava
  Majumdar. The MVAPICH Project: Evolution and Sustainability of an
  Open Source Production Quality MPI Library for
  HPC~\cite{Panda_WSSSPE}

\item Andy Terrel. Sustaining the Python Scientific Software
  Community~\cite{Terrel_WSSSPE}

\end{itemize}

\subsubsection*{Education \& Training}

\begin{itemize}

\item Ivan Girotto, Axel Kohlmeyer, David Grellscheid, Shawn
  T. Brown. Advanced Techniques for Scientific Programming and
  Collaborative Development of Open Source Software Packages at the
  International Centre for Theoretical Physics (ICTP)~\cite{Girotto_WSSSPE}

\item Thomas Crawford. On the Development of Sustainable Software for
  Computational Chemistry~\cite{Crawford_WSSSPE}

\end{itemize}

In addition, the following papers from other areas were also discussed
in this area.

\begin{itemize}

\item Charles R. Ferenbaugh, Experiments in Sustainable Software
  Practices for Future Architectures~\cite{Ferenbaugh_WSSSPE}

\item David Koop, Juliana Freiere, Cl\'{a}udio T. Silva, Enabling
  Reproducible Science with VisTrails~\cite{Koop_WSSSPE}

\item Sean Ahern, Eric Brugger, Brad Whitlock, Jeremy S. Meredith,
  Kathleen Biagas, Mark C. Miller, Hank Childs, VisIt: Experiences
  with Sustainable Software~\cite{Ahern_WSSSPE}

\item Sou-Cheng (Terrya) Choi. MINRES-QLP Pack and Reliable
  Reproducible Research via Staunch Scientific Software~\cite{Choi_WSSSPE}

\item Frank L\"{o}ffler, Steven R. Brandt, Gabrielle Allen and Erik
  Schnetter. Cactus: Issues for Sustainable Simulation
  Software~\cite{Loffler_WSSSPE}

\item Erik Trainer, Chalalai Chaihirunkarn, James Herbsleb. The Big
  Effects of Short-term Efforts: A Catalyst for Community Engagement
  in Scientific Software~\cite{Trainer_WSSSPE}

\end{itemize}

\newpage

\section{Attendees} \label{sec:attendees}  

The following is a partial list of attendees who were recorded on the
document~\cite{WSSSPE1-google-notes} that was being used for live note
taking at the workshop, or by the SC13 student volunteers, with some
additions also made by the authors of this report.

\begin{multicols}{3}
\setlength{\parindent}{0pt}

Jay Alameda

Gabrielle Allen

David Andrs

Brian Austin

Lorena A. Barba

David Bernholdt

Phil Bourne

Karl Broman

Sharon Broude Geva

Jed Brown

Maxine Brown

David Bruhwiler

Bruno Bzeznik

Alexandru Calotoiu

Jeffrey Carver

Shreyas Cholia

Peng Huat Chua

Neil Chue Hong 

John W. Cobb

Timothy Cockerill

Karen Cranston

Rion Dooley

Anshu Dubey

Marat Dukhan

Anne C. Elster

Ian Foster

Juliana Freire

Jeffrey Frey

Derek Gaston

Allison Gehrke

Brian Glendenning

Christian Godenschwager

Derek Groen

Edmund Hart

Magne Haveraaen

Steven Heaton

Oscar Hernandez

James Hetherington

Simon Hettrick

Jonathan Hicks

Kenneth Hoste

James Howison

Daisie Huang

Shao-Ching Huang

Tsutomu Ikegami

Kaxuya Ishimura

Christian Iwainsky

Craig Jackson

Wesley Jones

Randall Judd

Shusuke Kasamatsu

Daniel S. Katz

Kerk Kee

Kellie Kercher

Mads Kristensen

Carolyn Lauzon

Arnaud Legrand

Chris Lenhardt

Michael Levy

Frank L\"{o}ffler

Monica L\"{u}cke

Simon A. F. Lund

Arthur Maccabe

Paul Madden

Louis Maddox

Philip Maechling

Ketan Maheshwari

Brian Marker

Suresh Marru

Cezary Mazurek

James McClure

Matt McKenzie

Chris Mentzel

Paul Messina

Mike Mikailov

J. Yates Monteith

Reagan More

Rafael Morizawa

Pierre Neyron

Lucas Nussbaum

Patrick O'Leary

Manish Parashar

Cody Permann

Jack Perdue

John Peterson

Quan Pham

Marlon Pierce

Heather Piwowar

David Proctor

Sara Rambacher

Nicolas Renon

Jason Riedy

Todd Rimer

Bill Sacks

Andreas Schreiber

William Scullin

Andrew Slaughter

Jaraslaw Slawinski

Arfon Smith

Spencer Smith

James Spencer

Eric Stahlberg

Timothy Stitt

Hyoshin Sung

Fr\'{e}d\'{e}ric Suter

Shel Swenson

Yoshio Tanaka

Andy Terrel

George  Thiruvathukal

Keren Tomko

John Towns

Erik Trainer

Satori Tsuzuki

Matthew Turk

Eric van Wyk

Colin C. Venters

Brice Videau

Tajendra Vir Singh

Von Welch

Nancy Wilkins-Diehr

Theresa Windus

Felix Wolf

Rich Wolski

Lynn Zentner

\end{multicols}

\end{document}